\documentclass[]{jfm}
\usepackage{graphicx}
\usepackage{newtxtext}
\usepackage{newtxmath}
\usepackage{natbib}
\usepackage[colorlinks=true,linkcolor=blue,citecolor=blue,urlcolor=blue]{hyperref}
\usepackage[export]{adjustbox}
\usepackage{tikz}
\usepackage{wasysym}

\newcommand{\RomanNumeralCaps}[1]
\linenumbers

\newcommand{\fillcircb}{{\raisebox{.48ex}{\tiny$\CIRCLE$}}}
\newcommand{\reva}{}
\newcommand{\revb}{}
\newcommand{\revc}{}
\newcommand{\ours}{}

\title{Morphology of clean and surfactant-laden droplets in homogeneous isotropic turbulence}
\author{
Ianto Cannon,
Giovanni Soligo,
\and Marco E. Rosti  \corresp{\email{marco.rosti@oist.jp}}}
\affiliation{Complex Fluids and Flows Unit, Okinawa Institute of Science and Technology Graduate University, 1919-1 Tancha, Onna-son, Okinawa 904-0495, Japan
}
\begin{document}
\maketitle
\begin{abstract}
We perform direct numerical simulations of surfactant-laden droplets in homogeneous isotropic turbulence with Taylor Reynolds number $Re_\lambda\approx180$. The droplets are modelled using the volume of fluid method, and the soluble surfactant is transported using an advection-diffusion equation. Effects of surfactant on the droplet and local flow statistics are well approximated using a lower, averaged value of surface tension, thus allowing us to extend the framework developed by \citet{Hinze1955} and \citet{kolmogorov1949breakage} for surfactant-free bubbles to surfactant-laden droplets. 
\revc{We find that surfactant-induced tangential stresses play a minor role in this setup, thus allowing to extend the Kolmogorov-Hinze framework to surfactant-laden droplets.}
The Kolmogorov-Hinze scale \ours{$d_H$} is indeed found to be a pivotal length scale in the droplets' dynamics, separating the coalescence-dominated (droplets smaller than \ours{$d_H$}) and the breakage-dominated (droplets larger than \ours{$d_H$}) regimes in the droplet size distribution. We find that droplets smaller than \ours{$d_H$} have a rather compact, regular, spheroid-like shape, whereas droplets larger than \ours{$d_H$} have long, convoluted, filamentous shapes with a diameter equal to \ours{$d_H$}. This results in very different scaling laws for the interfacial area of the droplet. The \revb{normalised} area, \revb{$A/d_H^2$}, of droplets smaller than \ours{$d_H$} is proportional to \revb{$(d/d_H)^2$}, while the area of droplets larger than \ours{$d_H$} is proportional to \revb{$(d/d_H)^3$}, where $d$ is the droplet characteristic size. We further characterise the large filamentous droplets by computing the number of handles (loops of the dispersed phase extending into the carrier phase) and voids (regions of the carrier fluid completely enclosed by the dispersed phase) for each droplet. The number of handles per unit length of filament \ours{scales inversely with} surface tension. The number of voids is proportional to the droplet size and independent of surface tension. Handles are indeed an unstable feature of the interface and are destroyed by the restoring effect of surface tension, whereas voids can move freely in the interior of the droplets, unaffected by surface tension.
\end{abstract}

\begin{keywords}
surfactant, volume of fluid, Kolmogorov-Hinze scale, topology, Marangoni stresses, homogeneous isotropic turbulence, droplets
\end{keywords}

\section{Introduction}
\label{sec:intro}

\begin{figure}
  \centerline{
  \begin{tikzpicture}[x=6cm, y=6cm, font=\footnotesize]
    \node[anchor=south west,inner sep=0] (image) at (0,0) {\adjincludegraphics[width=6cm,trim={{.05\width} {.05\width} {.05\width} {.05\width}},clip]{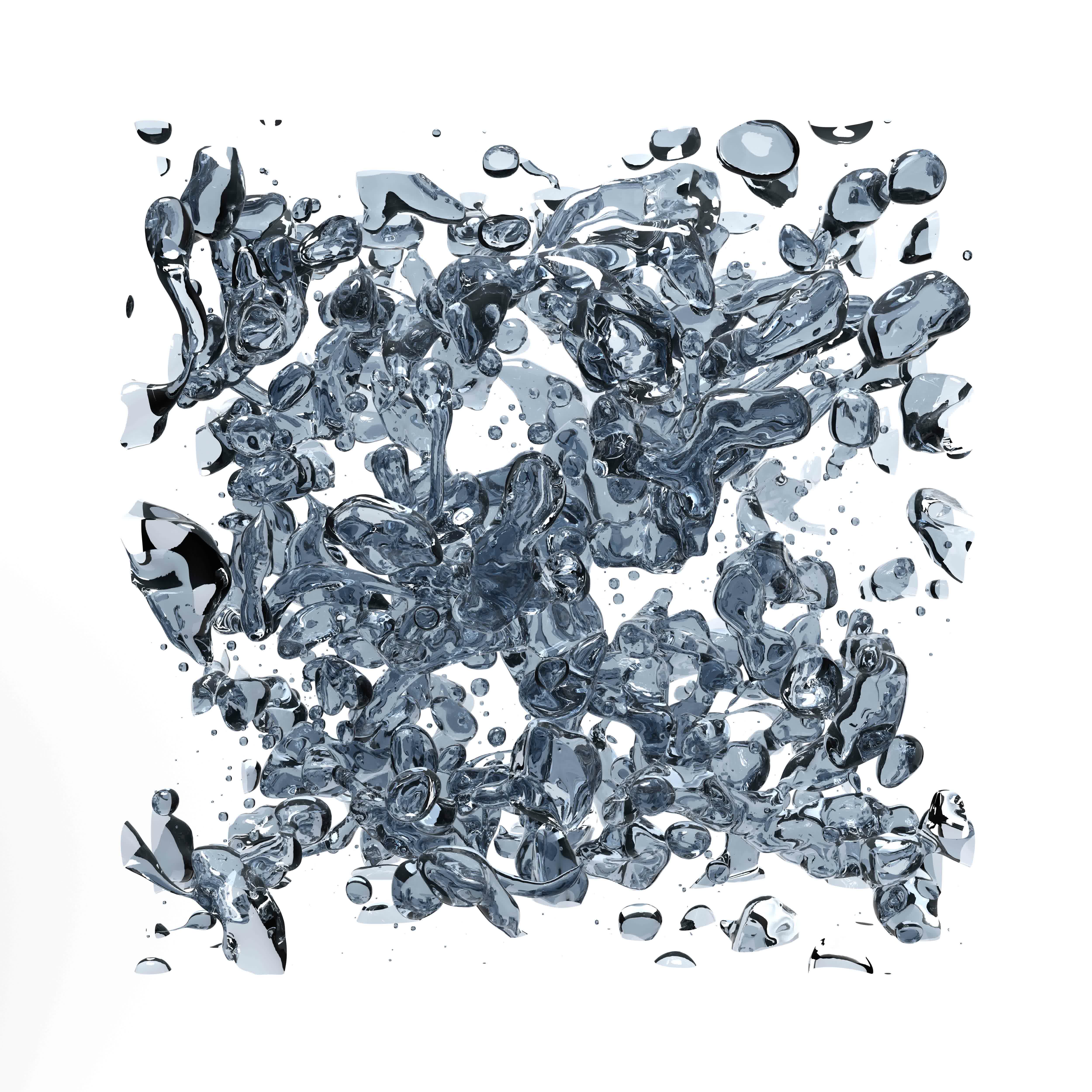}};
    \fill (image.south east) ++(-12mm,1mm) rectangle ++(.0489*147.5/186/.9, .5mm) node[anchor=west, shift={(0mm,0mm)}] {$d_H$};
  \end{tikzpicture}
  \begin{tikzpicture}[x=6cm, y=6cm, font=\footnotesize]
      \node[anchor=south west,inner sep=0] (image) at (0,0) {\adjincludegraphics[width=6cm,trim={{.05\width} {.05\width} {.05\width} {.05\width}},clip]{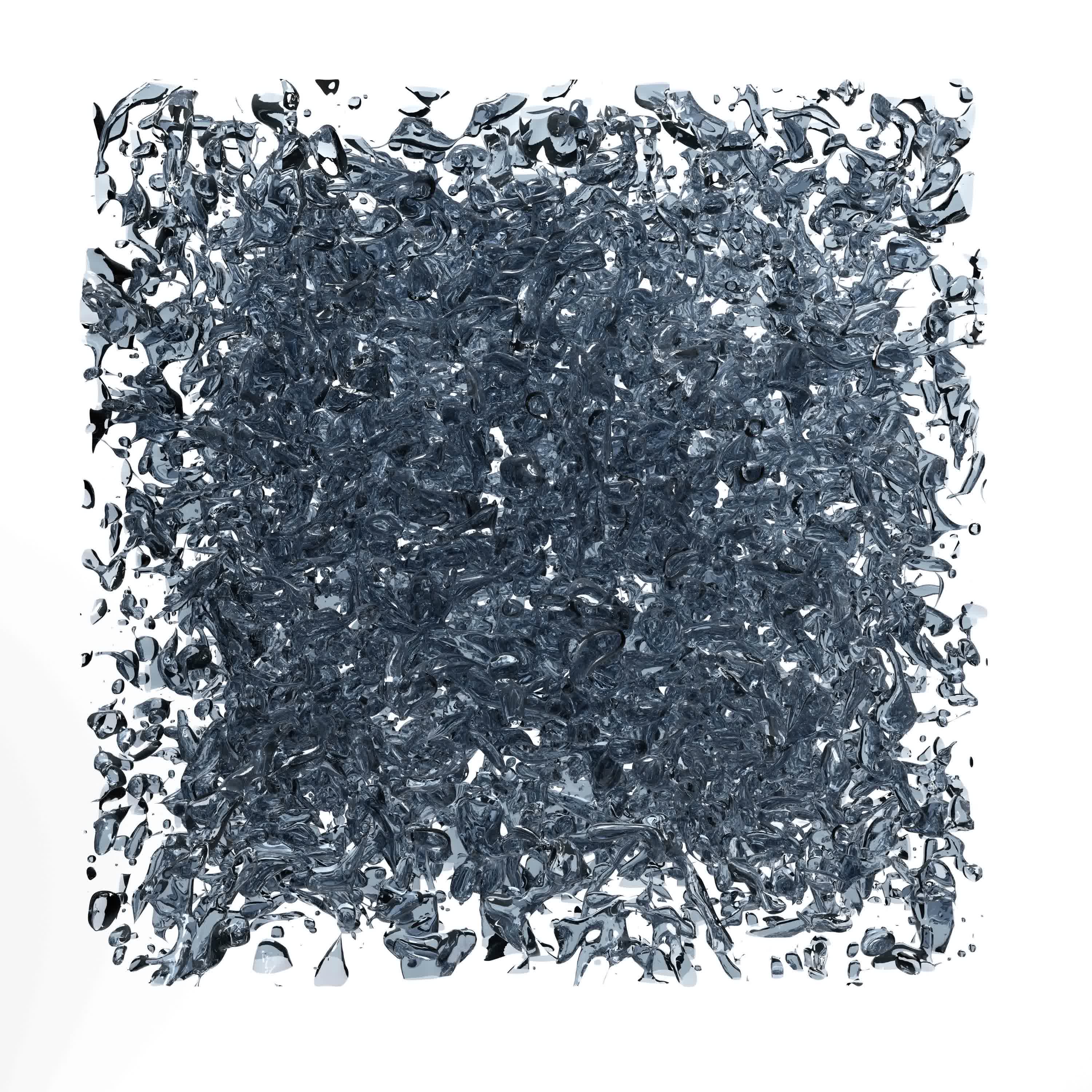}};
      \fill (image.south east) ++(-12mm,1mm) rectangle ++(.0232*147.5/186/.9, .5mm) 
      node[anchor=west, shift={(0mm,0mm)}] {$d_H$};
  \end{tikzpicture}  
  }
  \caption{Snapshots of the simulated domain in the cases (a) S05 and (b) S20 showing the interface of the droplets; the scale bar shows the Kolmogorov-Hinze scale for each case.}
\label{fig:quali}
\end{figure}

Droplet-laden turbulent flows are ubiquitous in nature and industry \citep{jahne1998air,karsa1999industrial,kralova2009,schramm20032,dickinson2010food}. A few examples include the capture of atmospheric CO$_2$ at the surface of seas and oceans, which is mediated by the entrainment of air bubbles by breaking waves \citep{merlivat1983gas,deane2002scale,pereira2018reduced}, or the dynamics of liquid jets and sprays which is of fundamental interest \ours{for} combustion, cooling, irrigation and firefighting. \citep{canu_where_2018,faeth_structure_1995,Herrmann2011,kooij_what_2018,mugele1951}. These flows rarely involve pure fluids: instead, they often include small amounts of impurities that may act as surface-active agents (surfactants). Surfactants are compounds that assemble at the fluid interface and modify the local surface tension. Even small amounts of surfactant can drastically change the flow behaviour, making their presence crucial in many practical scenarios \citep{koshy1988effect,Dobbs1989,sjoblom2005emulsions,takagi2011surfactant}.

The interface between the carrier phase and the dispersed phase, i.e. the droplets, serves as a conduit for various physical and chemical exchanges, such as heat, vapour \citep{scapin2020volume,onofre_ramos_dns_2022}, solutes, and aerosols \citep{de_leeuw_production_2011}. The rate of these exchanges is determined by the product of the interfacial flux and the interfacial area, emphasizing the pivotal role of interfacial characteristics. There has been considerable effort to estimate these exchanges via empirical correlations \citep{akita1974bubble,kelly1982interfacial,delhaye1994interfacial} or via population balance equations relying on the droplet size distribution and on droplet breakage and coalescence models \citep{LuoS_1996,BabinskyS_2002,andersson2006modeling,Andersson2006,martinez2010considerations,chan_formation_nodate,chan2021turbulent,gaylo2023fundamental}. \revc{Hence in this study, we aim to answer the question, \emph{``what is the interfacial area of surfactant laden droplets in turbulence?''}} Indeed, we \ours{measure} the interfacial area of each droplet and discover the presence of two universal regimes: the interfacial area of small droplets is proportional to the square of their characteristic size, whereas, for large droplets, it is proportional to the cube of their characteristic size. The information on the individual interfacial area, combined with the droplet size distribution, provides an estimate of the total interfacial area available. The two different regimes are directly linked to the shape of the droplets: small droplets are spheroid-like or ellipsoid-like, whereas large droplets take long, filamentous shapes. We find that the length scale separating these two regimes is the Kolmogorov-Hinze scale, defined as the maximum size of a droplet that is not broken apart by turbulent fluctuations; droplet breakage becomes prevalent for droplets larger than the Kolmogorov-Hinze scale. The concept of the Kolmogorov-Hinze scale originates from the works of \citet{kolmogorov1949breakage} and \cite{Hinze1955}, who applied \citeauthor{kolmogorov1941local}'s (1941) assumptions to droplets in turbulence. 
Some recent studies, however, have disputed the theoretical framework upon which Hinze's theory is based and hence the relevance of the Kolmogorov-Hinze scale: \citet{qi_fragmentation_2022} showed that droplets interact with eddies of a range of length scales, rather than solely with eddies of a size similar to the droplet, and \citet{vela-martin_memoryless_2022} showed that droplet breakup does, in fact, occur below the Kolmogorov-Hinze scale. \revc{Motivated by these studies, we ask, \emph{``does the Kolmogorov-Hinze framework hold in surfactant-laden flows?''} As we shall see, the answer is yes. That is,} our numerical simulations show that the Kolmogorov-Hinze scale can be used as a key parameter to describe the morphology of \ours{surfactant-laden} droplets in turbulence, and that the values obtained using Hinze's original formulation \citep{Hinze1955} are in good agreement with a more recent formulation, which uses the work done by the interface to define a length-scale for the droplet phase \citep{Crialesi-Esposito_Chibbaro_Brandt_2023}. Furthermore, we show that Hinze's theory can be extended to surfactant-laden droplets, provided that a suitable value of the surface tension is selected. In our setup, surfactant effects on the morphology of the droplets can indeed be well approximated using an averaged value of the surface tension, thereby maintaining the simplicity and efficacy of the Kolmogorov-Hinze framework.

Beyond an average reduction in surface tension, surfactants introduce more intricate dynamics into the flow: a surfactant is an additional phase that is transported by the local flow and by motion and deformation of the interface, and that reduces the value of surface tension according to its local concentration. This can lead to an inhomogeneous value of surface tension over the interface of the droplets, giving rise to Marangoni stresses, i.e. stresses that act tangentially to the interface and originate from surface tension gradients. Marangoni stresses have been shown to be crucial in hindering and preventing coalescence \citep{dai2008,Soligo2018_jcp}, reducing the rising velocity of bubbles \citep{takagi2011surfactant,Elghobashi2019}, and in the build-up of bubble layers in wall-bounded flows \citep{Tryggvason2008,takagi2011surfactant,tryggvason_direct_2015,LuMT_2017,ahmed_turbulent_2020}. An increase in the drag coefficient has been reported when adding surfactant to wall-bounded bubbly flows \citet{takagi2008effects,takagi2009surfactant,verschoof2016bubble}: surfactant reduces the size of the droplets, and causes a lower drag reduction compared to surfactant-free cases. Our study examines a statistically-stationary, homogeneous and isotropic multiphase flow at a moderate Reynolds number (see figure~\ref{fig:quali}). This type of flow does not allow for the build-up of large-scale surfactant gradients commonly found in up-flow and down-flow setups, where the velocity difference between the carrier and the dispersed phase generates and maintains a surfactant gradient along the interface of the droplets or bubbles \citep{takagi2008effects,LuMT_2017}. For this reason, we expect the effect of Marangoni stresses to be localized. \revc{In this study, we ask, \emph{``what is the effect of Marangoni stresses on droplet morphology in homogeneous-isotropic turbulence?''} As we shall show, surfactant effects on droplet morphology in our setup can be summarised as an average surface tension reduction, and the effect of Marangoni stresses can only} be appreciated by analysing the local flow dynamics at the interface.

To investigate the complex dynamics of clean and surfactant-laden droplets in turbulence we use direct numerical simulations. In recent years there has been a consistent growth in the number of numerical studies on multiphase turbulence, supported as well by the increased availability and capability of high-performance computing infrastructures. Multiphase turbulence is characterized by a wide range of scales, from the smallest, molecular-size interfacial scale, to the smallest flow scales -- the Kolmogorov length scale -- and up to the large-scale structures of the flow. The separation of scales usually spans over about eight to ten orders of magnitude, while direct numerical simulations on leading-edge high-performance computing systems can simulate about four orders of magnitude \citep{soligo2021turbulent}. The common choice is to simulate the larger scales of the multiphase flow, from the large-scale structures down to about the Kolmogorov length-scale, and introduce models for the smaller-scale physics \citep{tryggvason2010multiscale,tryggvason2013multiscale,soligo2021turbulent}. Several authors have devoted their attention to the development of models for the unresolved scales to be used in direct numerical simulations and large eddy simulations. In their original works \citet{kolmogorov1949breakage} and \citet{Hinze1955} identified the maximum size of a non-breaking droplet in turbulence. Experimental measurements \citep{deane2002scale,garrett2000} later showed that the existence of two different regimes in the droplet size distribution, separated by the Kolmogorov-Hinze scale. These works laid the foundations for the understanding of the dynamics of the droplets and for the development of sub-grid-scale models for the interfacial dynamics \citep{herrmann2013sub,xiao2014turbulent,evrard2019hybrid}. The power-law scaling exponents for the two regimes measured in experiments have been confirmed as well by numerical simulations: $-10/3$ for the breakage-dominated regime \citep{Perlekar2012,Skartlien2013,deike2016,mukherjee2019droplet,soligo2019breakage,RostiGJDB_2019,Crialesi-Esposito_Chibbaro_Brandt_2023} and $-3/2$ for the coalescence-dominated regime \citep{riviere_sub-hinze_2021,Crialesi-Esposito_Chibbaro_Brandt_2023}. The droplet size distribution and the population balance equation for the droplets are fundamental tools in the modelling of droplets of similar size to the grid resolution and smaller. \citet{Perlekar2012} correlated the instantaneous Weber number of droplets to their deformation, showing that large Weber numbers correspond to strongly deformed droplets. They simulated emulsions at increasing volume fractions and proved the validity of the Kolmogorov-Hinze scale at low volume fractions; they reported deviations from the Kolmogov-Hinze theory for more concentrated emulsions, possibly due to the effect of coalescence, which was neglected in the original works by Kolmogorov and Hinze. \citet{vela-martin_memoryless_2022} showed that drop-breakage is a memoryless process, i.e. the relaxation time of the droplet is much lower than its expected lifetime. In the same work, they debated the validity of the Kolmogorov-Hinze scale as an absolute threshold between breaking and non-breaking droplets, arguing that all droplets will eventually break apart, provided there is enough time for breakage. It was shown also that, in the absence of coalescence, the breakage rate depends on the Weber number alone. \citet{gaylo2023fundamental} investigated the fragmentation of bubbles in statistically-stationary homogeneous isotropic turbulence and characterised several fundamental time scales of the bubbles: the relaxation time, the expected lifetime and the time needed for the largest bubble to break down to the Kolmogorov-Hinze scale. It is now well known that the presence of a dispersed phase strongly modifies the dynamics of the flow at all scales: the interface extracts energy at the large flow scales and re-injects it into the flow at much smaller scales, competing with the classic turbulent energy cascade \citep{mukherjee2019droplet,perlekar_kinetic_2019}. This effect is reflected in a deviation from the $k^{-5/3}$ power-law scaling of the turbulent kinetic energy spectrum. When there is a considerable slip velocity between the droplet or bubble and the carrier fluid, a completely different scaling of the energy spectrum, $k^{-3}$, has been reported in numerical simulations \citep{roghair2011energy,pandey2020liquid,paul_role_2022} and experiments \citep{mercado2010bubble,prakash2016energy}. 

As part of our analysis of droplet morphology, we measure the Euler characteristic of the droplet interfaces. Similarly to the number of droplets in a flow, the Euler characteristic of an interface is an integer-valued topological invariant, and any change in its value requires a splitting or merging of interfaces. Despite its physical significance, the Euler characteristic has only recently been applied to multiphase flows: \citet{dumouchel_morphology_2022} linked the Euler characteristic to the Gaussian curvature of the droplets, and used it to parametrize the morphology of liquid droplets undergoing breakup. The Euler characteristic is commonly used in characterising the sintering of metal powders \citep{dehoff_experimental_1972,mendoza_evolution_2006}, classifying lung tissues \citep{boehm_automated_2008}, and correcting MRI scans of the human brain \citep{yotter_topological_2011}. In this article, we use the Euler characteristic to count the number of voids and handles on the droplets, demonstrating that the large droplets are made up of extremely interconnected filaments.

The article is structured as follows. We introduce the numerical method and the computational setup adopted for the simulations in section~\ref{sec:method}. Our findings are reported in section~\ref{sec:results}, where we first focus on the morphology of the dispersed phase, and then on the statistics of the local flow around the droplets. Finally, section~\ref{sec:concl} summarizes the main results presented in the present work. 

\section{Numerical model}
\label{sec:method}

We solve a system of equations including the momentum (\ref{eq:NS}) and mass (\ref{eq:mass}) conservation, the volume of fluid (\ref{eq:vof}) and surfactant (\ref{eq:surf}) transport equations to simulate the dynamics of an ensemble of breaking, coalescing and deforming finite-size droplets in homogenous isotropic turbulence. The two phases, the carrier fluid and the dispersed phase (i.e., the droplets) have the same density $\rho$ and dynamic viscosity $\mu$.
\begin{equation}
    \rho \frac{\partial \mathbf{u}}{\partial t} + \rho \mathbf{u}\cdot \nabla \mathbf{u} = -\nabla p +\nabla \cdot \left[ \mu \left( \nabla\mathbf{u}+\nabla \mathbf{u}^T\right) \right] + \nabla \cdot (\tau_c f_\sigma) + \mathbf{f_s},
    \label{eq:NS}
\end{equation}
\begin{equation}
    \nabla\cdot \mathbf{u}=0,
    \label{eq:mass}
\end{equation}
\begin{equation}
    \frac{\partial \phi}{\partial t} +\nabla \cdot(\mathbf{u}H)=0 ,
    \label{eq:vof}
\end{equation}
\begin{equation}
    \frac{\partial \psi}{\partial t} +\mathbf{u}\cdot \nabla \psi=\nabla \cdot (M_\psi \nabla \mu_\psi),
    \label{eq:surf}
\end{equation}
where $\mathbf{f_s}$ is the spectral forcing used to sustain turbulence. We use the one fluid approach, whereby the fluid velocity $\mathbf{u}$ and pressure $p$ are defined in both phases and continuous across the interface; the volume of fluid variable $\phi$ is used to define the instantaneous position of the interface. The volume of fluid can be understood as a colour function characterizing the local concentration of the dispersed phase: it is equal to $\phi=0$ in the carrier phase and to $\phi=1$ in the droplet phase. The volume of fluid method is an interface capturing method \citep{ProspTryg} where the concentration of each phase is transported using equation~\ref{eq:vof} and the interface is implicitly defined as the $\phi=0.5$ level. The effect of the interface on the flow is accounted for in the momentum equation via the surface tension forces: the Korteweg tensor $\tau_c=\mathcal{I}-\mathbf{n} \otimes \mathbf{n}$ \citep{KORTEWEG1901} accounts for the position and shape of the interface, and the surface tension equation of state $f_\sigma$ defines the local value of surface tension. In the definition of the Korteweg tensor, $\mathcal{I}$ is the identity matrix and $\mathbf{n}=-\nabla\phi/ ||\nabla \phi||$ is the unit-length, outward-pointing normal to the interface. The local surfactant concentration $\psi\in [0,1]$ is expressed as a fraction of the maximum surfactant concentration, which is usually determined by steric hindrance between surfactant molecules. Hence, $\psi$ is dimensionless in this formulation. To account for the effect of surfactant on surface tension, we use a modified Langmuir equation of state \citep{MuradogluT_2008,MuradogluT_2014,Soligo2018_jcp} $f_\sigma= \max[\sigma_{min},\sigma_0(1+\beta_s \log(1-\psi)) ]$, where $\sigma_0$ is the reference surface tension of a clean (i.e., surfactant-free interface), $\sigma_{min}$ the minimum surface tension, and $\beta_s$ the elasticity number. In the original formulation by \citet{bazhlekov2006numerical,Pawar1996}, the Langmuir equation of state provides a good fit at low to moderate surfactant concentration values; however, it fails to account for surfactant saturation dynamics at higher concentrations. Experimental measurements (reviewed by \citet{chang1995adsorption}) showed that, beyond a critical concentration of surfactant, surface tension no longer changes for increasing surfactant concentrations. Hence, to qualitatively account for surfactant saturation dynamics at the interface, we limit the surface tension to be greater than $\sigma_{min}$ at all points on the interface. In equation ~(\ref{eq:NS}), surface tension forces act perpendicular and tangential to the interface: a capillary component (normal to the interface) and a tangential component -- Marangoni stresses -- proportional to the surface tension gradient. The tangential component is characteristic of surfactant-laden flows, where surface tension changes along the interface according to the local surfactant concentration.

The volume of fluid $\phi$ is transported using a simple advection equation~(\ref{eq:vof}). The MTHINC (multi-dimensional tangent of hyperbola for interface capturing) method \citep{ii2012interface} is used to reconstruct the local volume of fluid value $\phi$ starting from the cell-local indicator function $H$. \revb{The transport equation for the cell-local indicator $H$ is integrated over a control volume, yielding equation~\ref{eq:vof} \citep{ii2012interface,RostiDB_2019,RostiGJDB_2019}.} To compute the surfactant chemical potential, we first calculate a signed-distance function $s$ and a smoothed colour function $\hat\phi$. A re-distancing equation is solved to compute the signed-distance function over the pseudo-time $\tau$
\begin{equation}
    \frac{\partial s}{\partial \tau}=\textnormal{sgn}(s_0) (1-|| \nabla s||),
    \label{eq:redist}
\end{equation}
where $\textnormal{sgn}$ is the sign function and the initial guess $s_0$ is found as $s_0=(2\phi-1)0.75\Delta$ \citep{albadawi2013influence}; this choice guarantees that the zero-level of the signed-distance function always corresponds to the interface \citep{russo2000remark,de_vita_effect_2019}. The signed-distance function is updated at every time iteration as the volume of fluid is advected. Next, we compute the smoothed colour function as $\hat\phi=\tanh{\frac{s}{3\Delta}}$, which is bounded in $-1\le \hat\phi \le 1$, and where the smoothing width is set to three times the grid spacing $\Delta$.  

An advection-diffusion equation~(\ref{eq:surf}), is solved to track the surfactant concentration $\psi$ in the \revb{entire} domain. We use a soluble surfactant: surfactant preferentially collects at the interface between the two fluids, but at the same time, it also dissolves in limited amounts in the bulk of the phases. \revb{We use a one-fluid formulation for the surfactant, in which the variable $\psi$ defines the surfactant concentration in the entire domain (i.e. in the bulk and at the interface).} 
The adsorption (accumulation of surfactant from the bulk to the interface), desorption (release of surfactant from the interface to the bulk) and diffusion dynamics of the surfactant phase are determined by the chemical potential of the surfactant. The chemical potential is made, in order, of three contributions: a free energy of mixing term, an adsorption term and a bulk-penalty term \citep{Engblom2013,Yun2014,Soligo2018_jcp},
\begin{equation}
    \mu_\psi=\alpha \ln \frac{\psi}{1-\psi}-\beta\frac{(1-\hat\phi^2)^2}{2}+\gamma\frac{\hat\phi^2}{2}. 
    \label{eq:chempot}
\end{equation}
The first term, the free energy of mixing term, favours a uniform surfactant distribution throughout the entire domain and plays the part of diffusion, with the coefficient $\alpha$ controlling the magnitude of the diffusive process. The adsorption term (second term) is a negative contribution to the free energy of the system because the accumulation of surfactant at the interface reduces the total energy of the surfactant configuration. The coefficient $\beta$ controls the adsorption dynamics. The last term, the bulk-penalty term, is representative of the cost of free surfactant, i.e., of surfactant dissolved in the bulk of the phases rather than adsorbed at the interface, and the coefficient $\gamma$ determines the energy cost of surfactant dissolved in the bulk phases. The adsorption term is maximum in magnitude at the interface ($\hat\phi=0$), indicating a decrease in the energy of the system, while the bulk-penalty term is maximum in the bulk of the phases ($\hat\phi=\pm1$) indicating an increase in the energy. The logarithmic formulation of the free energy of mixing term mandates for a non-constant mobility parameter $M_\psi=m \psi (1-\psi)$ \citep{Engblom2013}, with $m$ being a numerical coefficient controlling the magnitude of the diffusive-like surfactant dynamics. This choice of the mobility parameter ensures the boundedness of the surfactant concentration, $\psi \in [0,1]$.

Finally, we couple the volume of fluid method for simulating the interfacial dynamics with a phase-field-based method to track the concentration of a soluble surfactant. The volume of fluid method guarantees exact mass conservation of each phase and allows for a sharper interface compared to other diffuse-interface methods. We rely on a method to simulate surfactant dynamics that has been successfully adopted in the past to simulate flows with surfactant-laden interfaces \citep{Engblom2013,Yun2014,Soligo2018_jcp,soligo2019breakage,soligo2020deformation,soligo2020effect} using a phase field method to model interfacial dynamics. In particular, we employ a volume of fluid method to simulate the dynamics of the dispersed and carrier phases, and we use a smoothed colour function, $\hat\phi$, to couple the interfacial dynamics (based on the volume of fluid) with the surfactant dynamics (based on a phase field method). The smoothed volume of fluid field accounts for the interfacial dynamics while at the same time providing the diffuse-interface basis onto which the surfactant model is built. We thus combine the aforementioned strengths of the volume of fluid method with the advantages of a formulation of the surfactant phase that accounts for adsorption, desorption and diffusion in a thermodynamically consistent framework.

\subsection{Computational method}
The system of equations is discretized on a uniform Cartesian grid. The computational grid is staggered: pressure, density, viscosity, volume of fluid and surfactant concentration are defined at the cell centres, and the fluid velocities are stored at the cell faces. Spatial derivatives are approximated using a second-order finite difference scheme, and time advancement is performed via a second-order, explicit Adams-Bashforth scheme. A fractional-step method \citep{kim1985application} is adopted to advance the mass and momentum conservation equations in time, with the resulting Poisson equation for the pressure solved via a fast pressure solver. The volume of fluid is transported using a directional splitting method combined with an upwind scheme \citep{ii2012interface,RostiDB_2019}. The same scheme is used for the advective term in the surfactant transport equation. 

The surfactant is resolved on a refined grid to capture the steep concentration gradients at the interface and to keep a sharp interfacial profile. The smoothing width of the colour function $\hat \phi$ should be large enough to accurately discretize the surfactant profile across the interface, and at the same time, it should be small enough to keep the surfactant profile sharp. Using a refined grid thus allows us to capture the modelled interfacial dynamics while maintaining a thin interfacial surfactant layer. The finer computational grid used for the surfactant transport is still a staggered, uniform, Cartesian grid; linear interpolation is used to interpolate variables from/to the standard grid (for the velocity, pressure, density, viscosity and volume of fluid) to/from the fine grid (for the surfactant concentration). The surfactant transport is carried on the fine grid, with the velocity and smoothed volume of fluid fields interpolated to the fine grid. Surface tension forces are instead at first computed on the fine grid and then applied to the momentum conservation equation. 
Tests have been performed with different grid refinement factors: the pressure jump across the interface, the surfactant concentration value at the interface, and the total surfactant concentration show minimal changes compared to the reference case (i.e., unitary refinement factor). For the sake of comparison among the different cases, the smoothing width was kept constant in all cases, while in our numerical simulations, the smoothing width is adapted to the refinement factor, thus allowing for smaller values of the smoothing width and for a thinner surfactant interfacial layer.

We use the in-house code \textbf{\textit{Fujin}} to perform all the numerical simulations presented here. The code has been used and validated in the past on a variety of different flow configurations \citep{RostiGJDB_2019,olivieri2020dispersed, brizzolara2021fiber, cannon2021effect, mazzino2021unraveling, rosti2021shear, abdelgawad2023scaling,rosti2023large}. Further validation cases are available on the group's website, \url{https://groups.oist.jp/cffu/code}. Specific validation tests for the surfactant model and its implementation are reported in appendix~\ref{sec:validation}.

\subsection{Computational setup}
\label{sec:domain}

\begin{table}
\begin{center}
\def~{\hphantom{0}}
\begin{tabular}{rlccc|ccccccc}
case & &$\langle\phi\rangle$ & $\langle\psi\rangle$ & $We$ & $\eta/L$ & $\lambda/L$ & $Re_\lambda$ & $\langle\sigma\rangle_I/\sigma_0$ & $We_{e}$ & $d_H/L$ & $d_{H\sigma}/L$\\[3pt]
SP                        &                     \fillcircb      & 0   & 0      &         - &\textbf{8.39e-04}& \textbf{0.0220}&\textbf{178}&     -      &         -   &         -     &    -          \\
\color[HTML]{feb24c}C10   & \color[HTML]{feb24c}\fillcircb      & 0.1 & 0      &        10 &  8.49e-04       &     0.0224     &       180  &     1      &\textbf{9.91}&\textbf{0.0560}&\textbf{0.0565}\\
\color[HTML]{fc4e2a}{C20} & \color[HTML]{fc4e2a}\fillcircb      & 0.1 & 0      &        20 &  8.55e-04       &     0.0228     &       184  &     1      &      20.0   &    0.0373     &  0.0392       \\
\color[HTML]{bd0026}{C40} & \color[HTML]{bd0026}\fillcircb      & 0.1 & 0      &        40 &  8.58e-04       &     0.0231     &       188  &     1      &      39.3   &    0.0252     &  0.0290       \\ 
\color[HTML]{a6bddb}{S05} & \color[HTML]{a6bddb}$\blacklozenge$ & 0.1 & 0.1    &         5 &  8.48e-04       &     0.0226     &       183  &     0.40   &      12.7   &    0.0489     &  0.0370       \\
\color[HTML]{3690c0}{S10} & \color[HTML]{3690c0}$\blacklozenge$ & 0.1 & 0.1    &        10 &  8.71e-04       &     0.0242     &       199  &     0.40   &      25.8   &    0.0338     &  0.0317       \\
\color[HTML]{045a8d}{S20} & \color[HTML]{045a8d}$\blacklozenge$ & 0.1 & 0.1    &        20 &\textbf{8.85e-04}& \textbf{0.0247}&\textbf{201}&     0.41   &\textbf{49.5}&\textbf{0.0232}&\textbf{0.0241}\\
  \end{tabular}\\[9pt]
\caption{List of simulations performed. Here $\langle\cdot\rangle$ denotes an average over the domain volume, and $\langle\cdot\rangle_I$ denotes an average over the interface. All simulations have been carried out at constant volume fraction $\langle \phi \rangle$; an additional reference case (single phase, $\langle \phi \rangle=0$) is performed. We investigate four different Weber numbers ($We$) and two different values of the mean surfactant concentration $\langle \psi \rangle$. The measured values of the Kolmogorov length-scale $\eta$, the Taylor micro-scale $\lambda$, Taylor Reynolds number $Re_\lambda$, average surface tension $\langle \sigma \rangle_I $, effective Weber number $We_e$, and Kolmogorov-Hinze diameters $d_H$ and $d_{H\sigma}$ are reported. The largest and smallest values of each parameter are shown in bold. }
\label{tab:cases}
\end{center}
\end{table}

We perform direct numerical simulations in a cubic box of size $L$ with periodic boundary conditions in all spatial directions. 
Homogenous isotropic turbulence is sustained using the force $\mathbf{f_s}$ in equation~\ref{eq:NS}. We use the spectral forcing scheme proposed by \cite{eswaran_examination_1988}, whereby the flow is forced in a shell of Fourier modes ${2\pi/L_{a} \leq |\mathbf{k}| \leq 2\pi/L_{b}}$, and the force on each mode evolves randomly in time \citep{Uhlenbeck_Ornstein_1930} with variance $\rho\sigma_L^2$ and relaxation time $T_L$. 
Hence, ${U_L\equiv\sigma_L^{2/3}T_L^{1/3}(L/2\pi)^{1/3}}$ is the characteristic velocity scale of the forcing. We set the forcing Reynolds number ${Re_L\equiv \rho U_L L/(2 \pi\mu)=41.6}$ to give a turbulence intensity that is tractable on our numerical grid. We choose the dimensionless relaxation time ${T_L^*\equiv 2\pi T_L U_L/L=2.08}$ to give variations at the timescale of the large eddy turnover time. To prevent droplets from spanning the entire periodic domain, we force at a length-scale smaller than $L$~\citep{mukherjee2019droplet,Crialesi-Esposito_Chibbaro_Brandt_2023}. Hence, the minimum and maximum wavelengths of forcing are set to $L_{b}=L/3$ and $L_{a}=L/2$, respectively. 

The computational domain is discretized using an equispaced Cartesian grid with $N=500$ grid points in all directions; to better resolve the sharp surfactant gradients at the interface and keep a sharper surfactant profile across the interface, the surfactant transport equation is resolved on a twice-refined grid, with $N_\psi=1000$ grid points in all directions. A refinement factor of 2 for the surfactant concentration grid has been selected as it significantly improves the surfactant profile's sharpness at the interface while keeping the computational cost within reasonable limits. With this refinement factor, the computational cost increases by roughly $\sim 25\%$ and the storage requirements by $\sim 116\%$.

We report in table~\ref{tab:cases} the chosen parameters for all cases reported in this article. We use one single-phase reference case ($\langle \phi \rangle=0$), three cases with clean droplets ($\langle \psi \rangle=0$), and three cases with surfactant-laden droplets ($\langle \psi \rangle=0.1$). The droplet-laden flows were initialised using fluid velocity and pressure from the single-phase reference case once it had reached a statistically steady state. A single spherical droplet of radius $R\simeq0.288L$ (corresponding to $\langle \phi \rangle=0.1$) was initialized at the centre of the computational box. Due to the action of the surrounding turbulent flow, the droplet deforms and breaks apart into smaller droplets. For our surfactant-laden cases, the surfactant is initially distributed in the domain following the equilibrium profile with $\psi=0.1$ in the bulk phase, computed by zeroing the gradient of the chemical potential. 
We focus on a highly-soluble surfactant and set the coefficients of the surfactant chemical potential to $\alpha={0.0242 u'_0}^2$, $\beta={0.0121 u'_0}^2$ and $\gamma={0.0121 u'_0}^2$, and the numerical coefficient of the mobility parameter to $m=0.0307L_b/u'_0$. For the flows with droplets, we fix the reference surface tension $\sigma_0$, allowing us to define a reference Weber number $We\equiv\rho {u'_0}^2 L_{b}/\sigma_0$ based on the single phase root mean square velocity $u'_0$ and the minimum wavelength of the forcing $L_{b}$. We select a moderate-strength surfactant with elasticity number $\beta_s=5$ and a relatively high surfactant saturation concentration, yielding a low minimum surface tension, $\sigma_{min}=0.1\sigma_0$. 

\ours{To verify the grid-independence of the present results, we performed two additional simulations on a more refined grid, $N_f=2N=1000$ grid points: a single phase (same parameters as case SP) case and a surfactant-free multiphase case (same parameters as case C10). Results on the fine grid, $N_f=1000$, are reported in the following and summarized here. The scale-by-scale energy budget confirms that the standard grid, $N=500$, is sufficient to capture all relevant turbulence scales. For the multiphase case, we do not report any relevant differences between the statistics of the fine grid and the standard grid cases; we do observe the presence of smaller droplets in the fine grid case, as expected due to the increased resolution.}
\revb{Clearly, when adopting a more refined grid in the numerical simulation of multiphase flows, we do expect differences in the morphology of the dispersed phase, as smaller droplets and interfacial features (for instance, ligaments and sheets) can be resolved. This impacts the droplet size distribution and the simulation of interface breaking and merging. 
Interface breaking is a relatively fast phenomenon that can be well approximated using a continuum formulation \citep{eggers1995,eggers1997nonlinear,soligo2019breakage}. It has been shown that grid resolution has a minor effect on the simulation of interface breaking \citep{Herrmann2011,Lu2018,Lu2019}.
Interface merging, on the other hand, is dependent on the grid resolution: the final stages of interface merging depend on physics acting at scales as small as the molecular scale \citep{aarts2004direct,aarts2005hydrodynamics,Chen2004,kamp2017drop,perumanath2019,mackay1963gravity}. These small scales are not resolved in continuum simulations of multiphase flows and interface merging occurs whenever the interface-interface distance becomes smaller than the grid spacing. Since coalescence occurs based on an artificial length scale, it has been named numerical coalescence. 
As interface merging involves physics down to the molecular scale, a complete simulation of coalescence cannot be attained solely by grid refinement \citep{Scardovelli1999,tryggvason2013multiscale,soligo2019breakage,soligo2021turbulent}. 
Numerical coalescence is still an open issue and several approaches and models have been put forward to improve the simulation of interface merging, see \citet{soligo2021turbulent} for a general review. In our case, we do not employ any models for interface merging, and as shown in figures~\ref{fig:dsd},~\ref{fig:deformation},~\ref{fig:kurv},~\ref{fig:area}, and~\ref{fig:genus}, grid resolution effects are seen only for statistics on the smallest droplets.}

\section{Results}
\label{sec:results}

In table~\ref{tab:cases}, we report integral quantities from all cases studied. Length scales of the turbulent flow are the Kolmogorov scale $\eta\equiv(\mu/\rho)^{3/4}\varepsilon^{1/4}$, and the Taylor micro scale $\lambda\equiv\sqrt{15\mu/(\rho\varepsilon)}u'$, where $u'$ is the root-mean-square velocity of the flow and $\varepsilon$ is the mean dissipation rate. The Taylor Reynolds number $Re_\lambda\equiv \rho u' \lambda / \mu$ is 178 in the single-phase case, and as was previously observed by \citet{RostiGJDB_2019} and \citet{crialesi-esposito_modulation_2022}, $Re_\lambda$ increases slightly when droplets are present. 
The surface tension averaged over the interface~$\langle \sigma \rangle_I$, is the same as the reference surface tension~$\sigma_0$ in the cases with clean droplets. However, it is reduced by more than half in the presence of surfactant. This motivates us to define an effective Weber number, $We_{e}\equiv \rho u'^2 L_{b}/\langle \sigma \rangle_I$ to better compare the different cases. 

The Kolmogorov-Hinze diameter $d_H\equiv0.725 \langle \sigma \rangle_I^{3/5}\rho^{-3/5} \varepsilon^{-2/5}$ is an estimate of the diameter of the largest droplet which does not break up. It is made by balancing surface tension with turbulent velocity fluctuations, using an empirical constant of $0.725$ \citep{Hinze1955}. We also use a more recent formulation of the Kolmogorov-Hinze diameter from \cite{Crialesi-Esposito_Chibbaro_Brandt_2023}; at large scales, droplets predominantly break up, and the interface takes energy from the flow (negative work), whereas at smaller scales, droplets predominantly coalesce and the interface returns energy to the flow (positive work). The length scale at which the work done by the interface is zero is defined as $d_{H\sigma}$. The two estimates of the Kolmogorov-Hinze diameter are in fairly good agreement in the cases with and without surfactant.

\begin{figure}
  \centerline{
  \begin{tikzpicture}
    \node at (0,0) {\includegraphics[width=.55\textwidth]{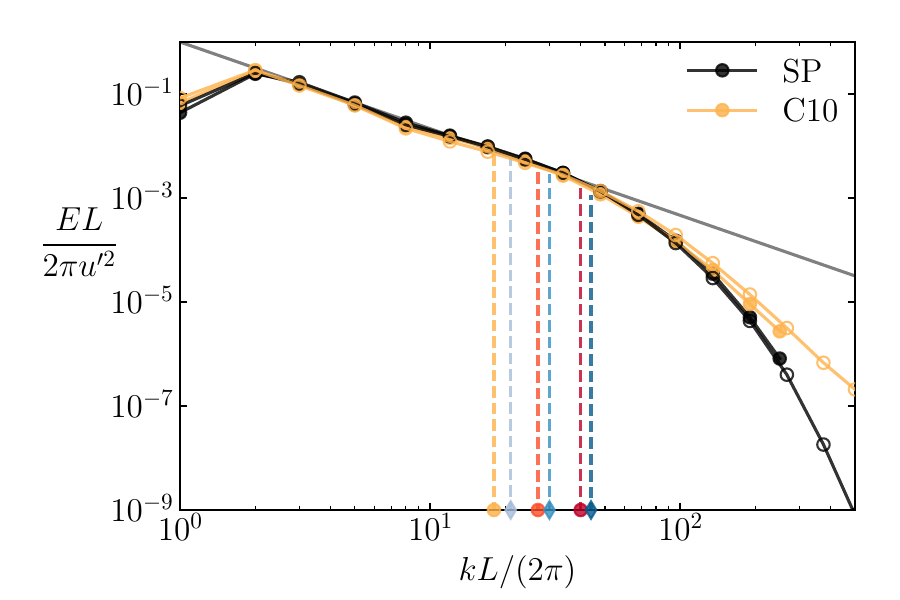}};
    \node at (7,0) {\includegraphics[width=.55\textwidth]{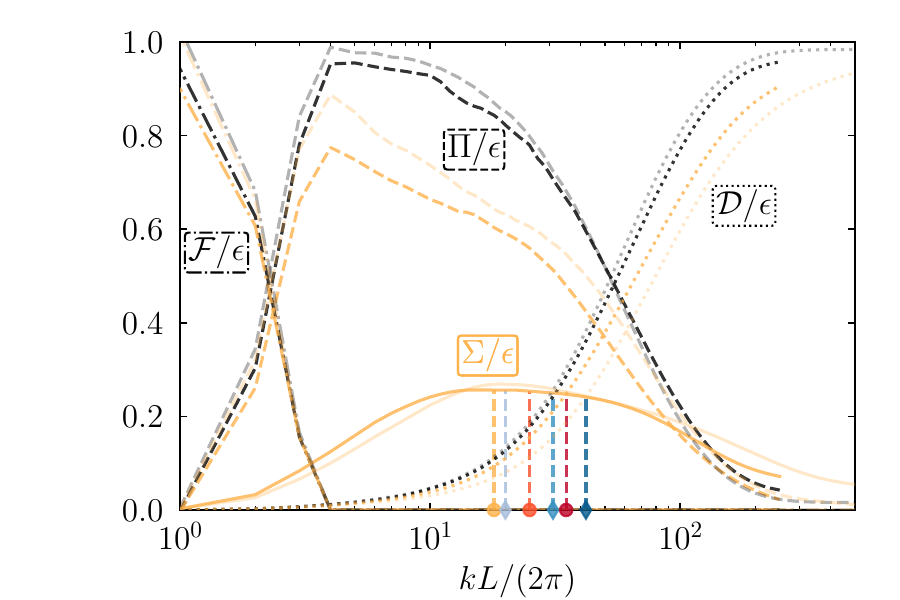}};
    \node at (-3.0,1.8){$(a)$};
    \node at (+4.0,1.8){$(b)$};
  \end{tikzpicture}
  }
  \caption{$(a)$ Turbulent kinetic energy spectrum $E$ for cases SP and C10, made dimensionless using the domain size $L$ and the root mean square velocity $u'$ of each case. Vertical dashed lines show the wavenumber of the Kolmogorov-Hinze scale $k_H\equiv 2\pi/d_H$ for all the cases with droplets. The solid grey line above reports the Kolmogorov scaling~$k^{-5/3}$. \ours{Data from simulations performed on a fine grid, $N_f=1000$, are reported with empty markers.} $(b)$~Scale-by-scale energy balance for cases SP and C10. Energy flux due to forcing $\mathcal{F}$, viscous dissipation $\mathcal{D}$, advection $\Pi$, and surface tension $\Sigma$ are plotted using dot-dashed, dotted, dashed, and solid lines, respectively\ours{; semi-transparent lines identify the data from the simulations performed on a fine grid}. Vertical dashed lines mark $k_{H\sigma}\equiv 2\pi/d_{H\sigma}$, the wavenumber at which $\Sigma$ is maximum for every droplet case.}
\label{fig:spec}
\end{figure}
Figure~\ref{fig:spec}a shows the kinetic energy spectra of the turbulent flows. In all cases, the most energetic modes are in the range $2 \le k L/(2\pi)\le3$, where turbulent forcing is applied. The single-phase case shows the $k^{-5/3}$ scaling predicted by \cite{kolmogorov1941local}, persisting for over a decade of wavenumbers. As it has been previously reported by \cite{perlekar_kinetic_2019,RostiGJDB_2019} and \cite{crialesi-esposito_modulation_2022}, the cases with droplets show a slight reduction in energy at small wavenumbers, and an increase at large wavenumbers. We also see from figure~\ref{fig:spec} that the Kolmogorov-Hinze scale is well within the inertial range. Hence, we can assume that self-similarity applies to the turbulent velocity fluctuations which dictate droplet deformation and breakup, investigated in the following subsection.
\ours{The turbulent kinetic energy spectrum computed on the fine grid ($N_f=1000$, shown with empty markers) superposes onto that computed on the standard grid and further extends into the small scales, thus indicating that the standard grid is sufficiently refined to simulate turbulence accurately.}
Our simulations are in a statistically steady state, and so the fluid kinetic energy contained in each wavenumber is constant in time, and the energy flux through each wavenumber $k$ is constant and equal to the energy injection rate $\epsilon$. This is expressed by the equation ${\mathcal{F}(k)+\Pi(k)+\mathcal{D}(k)+\Sigma(k)=\epsilon}$, where the terms on the left-hand side are the energy flux due to forcing, advection, viscous dissipation, and surface tension, respectively. We calculate these terms using the method given in the supplementary information of \citet{abdelgawad2023scaling} (see also chapter 6 of \citet{POPE2000}). Namely, we take a three-dimensional Fourier transform of each term in the Navier-Stokes equation~(\ref{eq:NS}), multiply by the fluid velocity, and integrate over the region bounded by a sphere of radius $k$ in wavenumber space. For dissipation, we choose the region inside the sphere; for the other terms, we choose the region outside the sphere. This way, as $k \to \infty$, $\mathcal{D}=\epsilon$ and $\mathcal{F}=\Pi=\Sigma=0$.
Figure~\ref{fig:spec}b shows the energy balance for the single-phase flow and a droplet-laden flow. The single-phase case shows the canonical Richardson cascade; energy is injected by forcing at the large scales and carried to smaller scales by advection, where it is dissipated by viscosity. In the droplet-laden case, surface tension also carries energy to smaller scales.
\ours{Minor differences can be appreciated between the standard grid cases (solid lines) and the fine grid cases (semi-transparent lines). For the single phase case, all the energy is almost completely transferred into dissipation and a plateau in the viscous dissipation is reached at large wavenumbers. A similar result is also observed for the multiphase case, where a fraction of the energy remains stored at small scales in the surface tension term for both grid resolutions.}

\subsection{Droplet statistics} 
\begin{figure}
  \centerline{
  \begin{tikzpicture}
    \node at (0,0) {\includegraphics[width=.55\textwidth]{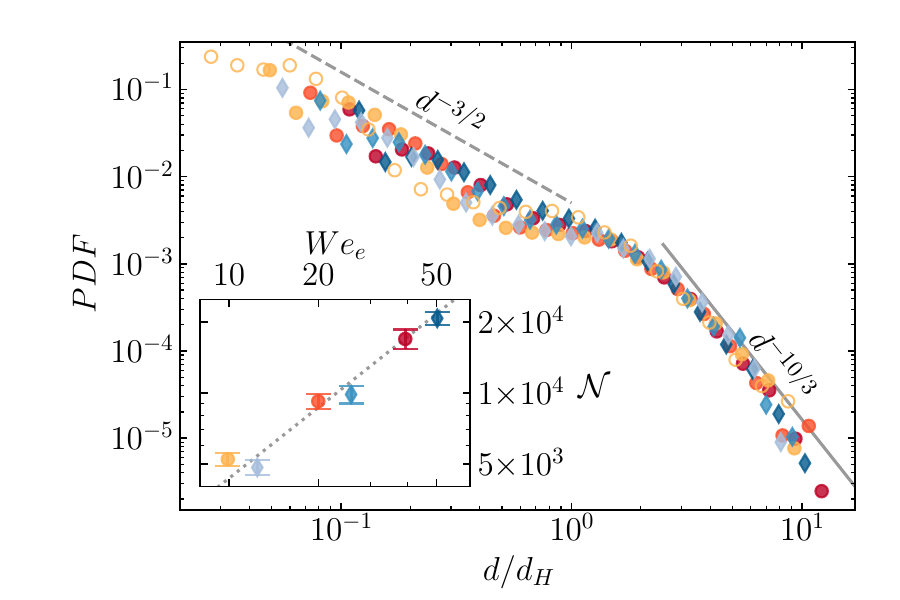}};
    \node at (7,0) {\includegraphics[width=.55\textwidth]{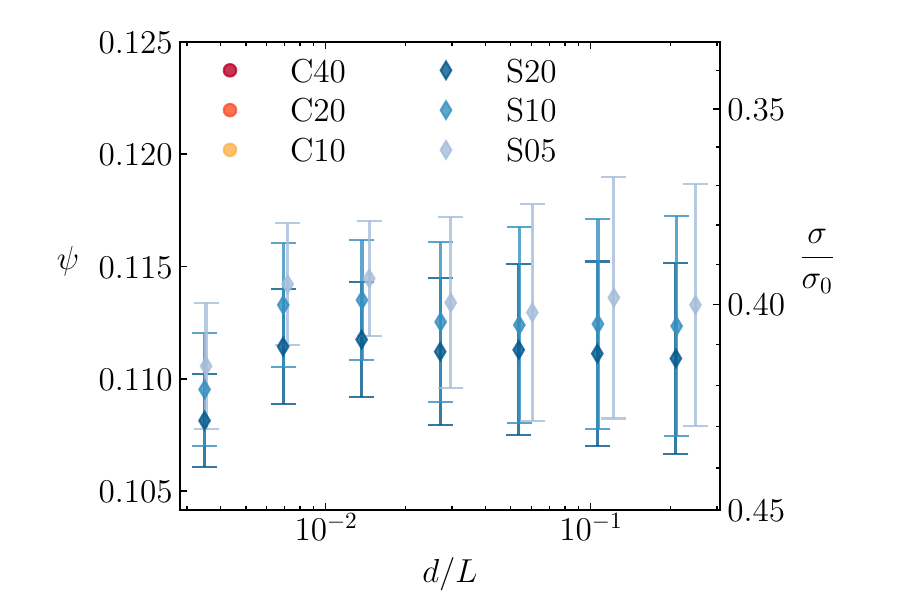}};
    \node at (-3.0,1.8){$(a)$};
    \node at (+3.8,1.8){$(b)$};
  \end{tikzpicture}
  }
  \caption{$(a)$ PDF of the droplet diameters $d$, with the dashed and solid lines showing scalings previously found in the coalescence and breakup regimes \citep{deane2002scale}, respectively; \ours{data from the simulation performed on a fine grid, $N_f=1000$, is reported with empty markers}. The inset reports the total number of droplets $\mathcal{N}$ 
  in each case, with error bars showing the root mean square variation in time. The dotted line is a fit ${\mathcal{N}=434We_e}$. In $(b)$, we calculate the mean and standard deviation of the surfactant concentration $\psi$ on the interface of each drop; we then average these values over the ensemble of drops of the same size. The right-hand axis shows the normalized interfacial surface tension resulting from the presence of surfactant. 
  }
\label{fig:dsd}
\end{figure}

The inset of figure~\ref{fig:dsd}a shows the average number of droplets $\mathcal{N}$ in our simulations. To identify and count each droplet, we use a stack-based, six-way flood-fill on the computational cells characterized by ${\phi\ge 0.5}$; the algorithm is a direct extension to three-dimensional space of the two-dimensional four-way flood fill algorithm \citep{newman1979principles}. 
The number of droplets has been counted over several instantaneous snapshots and averaged in time once the simulation has reached a statistically steady state, i.e. once the Taylor-Reynolds number and the number of droplets fluctuate about a constant mean value. 
We note that clean and surfactant-laden cases at similar values of the effective Weber number, i.e. S05 \& C10, S10 \& C20, S20 \& C40, have approximately the same average number of droplets. 
This result suggests that the effect of surfactant on the dispersed phase manifests mainly as an average surface tension reduction, with negligible effects from Marangoni stresses. As previously found by \citet{RostiGJDB_2019} in shear turbulence, we see that the number of droplets is proportional to the Weber number, in our case with a factor $\mathcal{N}=434We_e$.

The average number of droplets $\mathcal{N}$ is of the order $10^4$. This large sample size allows us to make accurate statistics of the droplets, even when binned by their equivalent diameter~$d$. We define the equivalent diameter of a droplet 
\begin{equation}
d\equiv(6V/\pi)^{1/3},
\label{eq:diam}
\end{equation}
as the diameter of a sphere with volume $V$, where $V$ is the volume of said droplet. The droplet size distribution for all cases is shown in figure~\ref{fig:dsd}(a). We observe a collapse of all the curves when the droplet size $d$ is normalized by the Kolmogorov-Hinze scale in each case. Note that the Kolmogorov-Hinze scale separates two different regimes: the coalescence-dominated regime for $d/d_H<1$ and the breakage-dominated regime for $d/d_H>1$. The Kolmogorov-Hinze scale \citep{Hinze1955} is defined as the largest droplet that resists breakage due to turbulent fluctuations. The coalescence-dominated regime characterizes droplets smaller than the Kolmogorov-Hinze scale; breakage is highly unlikely for these droplets, which instead are in a state of constant coalescence. On the other hand, droplets larger than the Kolmogorov-Hinze scale are prone to breaking apart. The droplet size distribution shows a clear power-law behaviour in the breakage- and coalescence-dominated regimes. Using dimensional arguments the exponents for the two regimes have been obtained: $-3/2$ for the coalescence-dominated regime \citep{deane2002scale} and $-10/3$ for the breakage-dominated regime \citep{garrett2000}. \citet{deane2002scale} measured the bubble size distribution in breaking waves and found good agreement between the experimental measurements and the analytical scalings. Several previous computational works confirmed the same $-10/3$ power-law exponent in the breakage-dominated regime \citep{Skartlien2013,deike2016,mukherjee2019droplet,soligo2019breakage,RostiGJDB_2019,Crialesi-Esposito_Chibbaro_Brandt_2023}, while fewer works have captured the $-3/2$ power-law exponent for the coalescence-dominated regime \citep{riviere_sub-hinze_2021,Crialesi-Esposito_Chibbaro_Brandt_2023}. Our results in figure~\ref{fig:dsd}(a) show \revc{a size distribution which is compatible with the breakage-dominated regime scaling, $-10/3$. However, the available data in the breakage-dominated regime spans at most one decade; hence, inferring an accurate power-law scaling here is challenging.}
For the low $We_e$ cases, we see some deviation from the $-3/2$ scaling just below the Kolmogorov-Hinze scale. However, further into the coalescence-dominated regime, around $d/d_H\approx10^{-1}$, all cases follow the $-3/2$ scaling. Furthermore, we show that the macroscopic effect of surfactant on the droplet size distribution is well captured by considering a lower, average surface tension value when computing the Kolmogorov-Hinze scale. A similar result was previously obtained for surfactant-laden flows \citep{Skartlien2013,soligo2019breakage}, although for the breakage-dominated regime alone. Here, we extend the result to droplets smaller than the Kolmogorov-Hinze scale. 
\ours{We also report the droplet size distribution data for the case C10 on a more refined grid (empty markers); a good agreement is observed with data from the same case on the standard grid. Clearly, on the more refined grid, we are able to capture smaller droplets; hence, the droplet size distribution extends slightly further to smaller diameters.}

Figure~\ref{fig:dsd}b shows the dependence of surfactant concentration $\psi$ on the equivalent droplet diameter $d$. For these values and error bars, the average surfactant concentration and its standard deviation were computed at the interface of each droplet, and then averaged over all the droplets of size $d$. This way, the error bars capture not the variation of $\psi$ between droplets, but the average variation on each droplet, which governs Marangoni stresses. 
We see that the average surfactant concentration at the interface is higher than the initial concentration in the bulk phase ($\psi=0.1$), as from equation~\ref{eq:chempot} it is energetically favourable for the surfactant to assemble on the interface. We observe a trend in the average surface tension value at the interface for increasing values of the Weber number: as the total amount of interfacial area increases (the total number of droplets is roughly proportional to the Weber number), the average surfactant concentration at the interface reduces. For the surfactant parameters considered in this study, there is little dependence of the mean surfactant concentration at the interface on the size of the droplets. The average surfactant concentration for all droplet sizes is about $\psi\approx 0.115$, with a slightly lower value for the smallest droplets. This results in an average reduction of the surface tension to approximately 40\% of its clean value. We note that a surface tension reduction of around one-half is typical of real-world surfactant-laden interfaces, such as Tween 80 and NaCl in water \citep{qazi_dynamic_2020}. Error bars show the mean standard deviation of surfactant concentration on a droplet with equivalent diameter $d$. The standard deviation on the droplets has instead a mild dependence on the characteristic size of the droplet, showing about a twofold increase between the smallest and the largest droplets (i.e., over a $\sim100\times$ increase in the droplet equivalent diameter). The variation in surface tension is approximately 15\% of the mean surfactant concentration at the interface, corresponding to about 8\% change in surface tension on each droplet. 

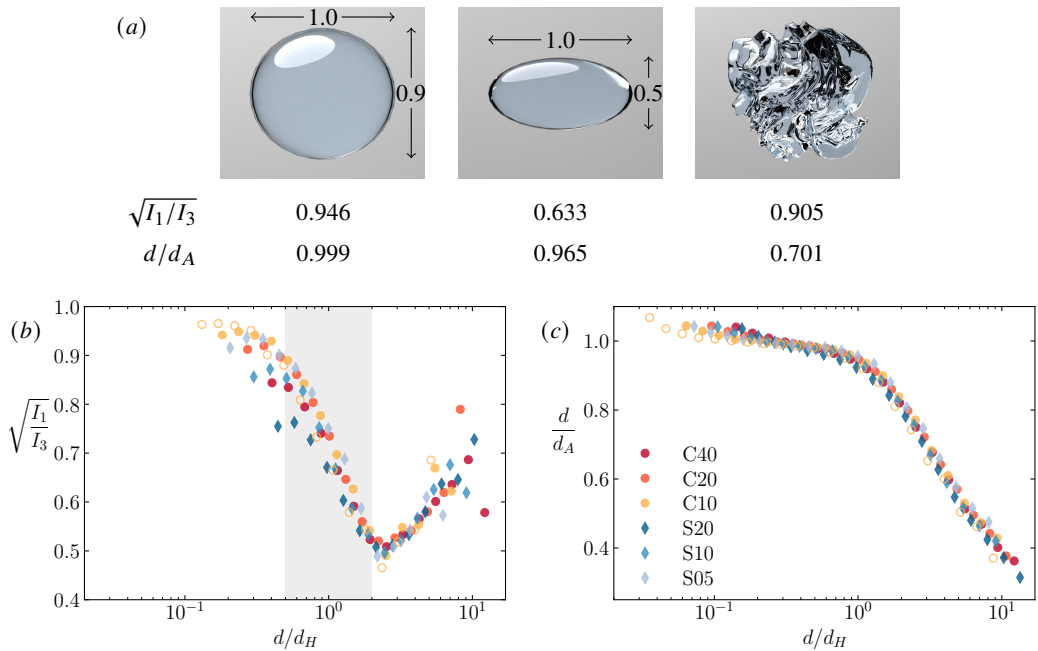
\begin{figure}   
    \centerline{
    \begin{tikzpicture}
        \node at (3.5,4.4) {
            \begin{tabular}{rccc}
            &\begin{tikzpicture}
                \node (eli) {\adjincludegraphics[width=0.2\textwidth,trim={{.2\width} {.25\width} {.2\width} {.25\width}},clip]{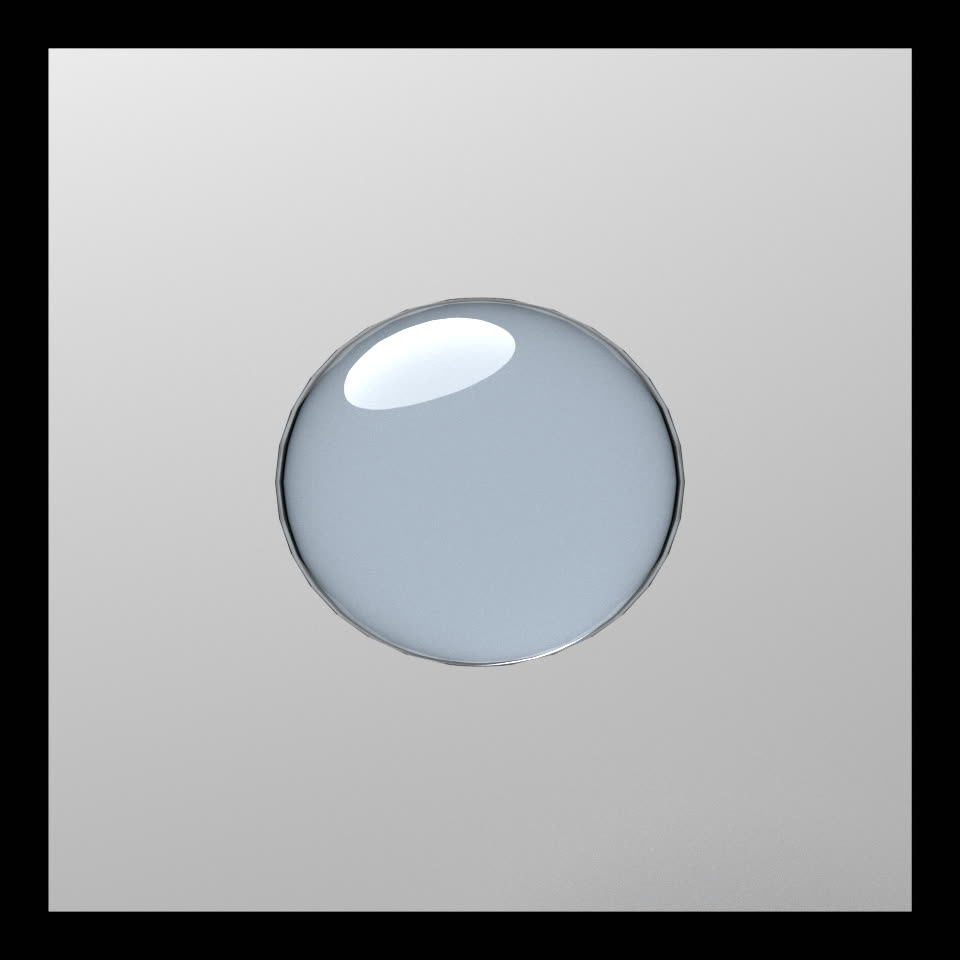} };
                \draw[->] (-.25,1) -- (-.95,1);
                \draw (0,1) node {1.0};
                \draw[->] (.25,1) -- (.95,1);
    
                \draw[->] (1.17,-.2) -- (1.17,-.86);
                \draw (1.17,0) node {0.9};
                \draw[->] (1.17,.2) -- (1.17,.86);            
            \end{tikzpicture}&
            \begin{tikzpicture}
                \node (eli) {\adjincludegraphics[width=0.2\textwidth,trim={{.2\width} {.25\width} {.2\width} {.25\width}},clip]{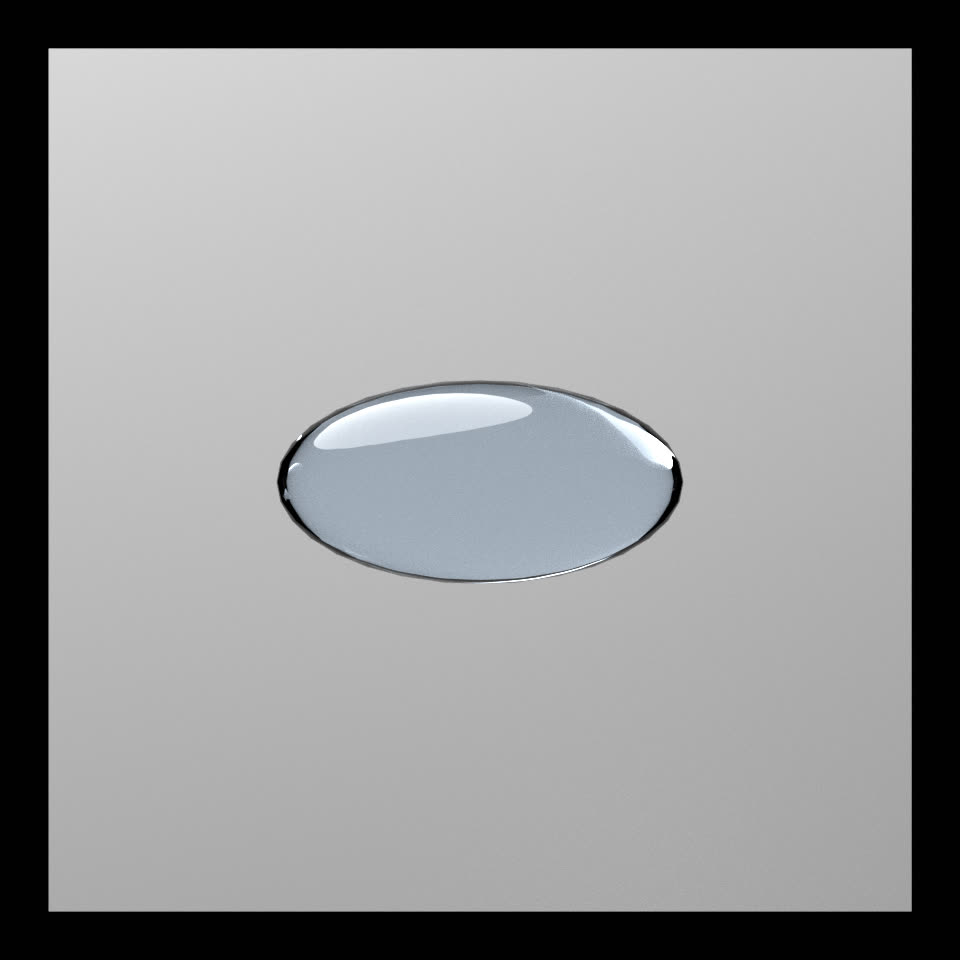} };
    
                \draw[->] (-.25,.7) -- (-.95,.7);
                \draw (0,.7) node {1.0};
                \draw[->] (.25,.7) -- (.95,.7);
    
                \draw[->] (1.17,-.2) -- (1.17,-.48);
                \draw (1.17,0) node {0.5};
                \draw[->] (1.17,.2) -- (1.17,.48);  
           
            \end{tikzpicture}&
            \begin{tikzpicture}
                \node (img) {\adjincludegraphics[width=0.2\textwidth,trim={{.055\width} {.129\width} {.055\width} {.129\width}},clip]{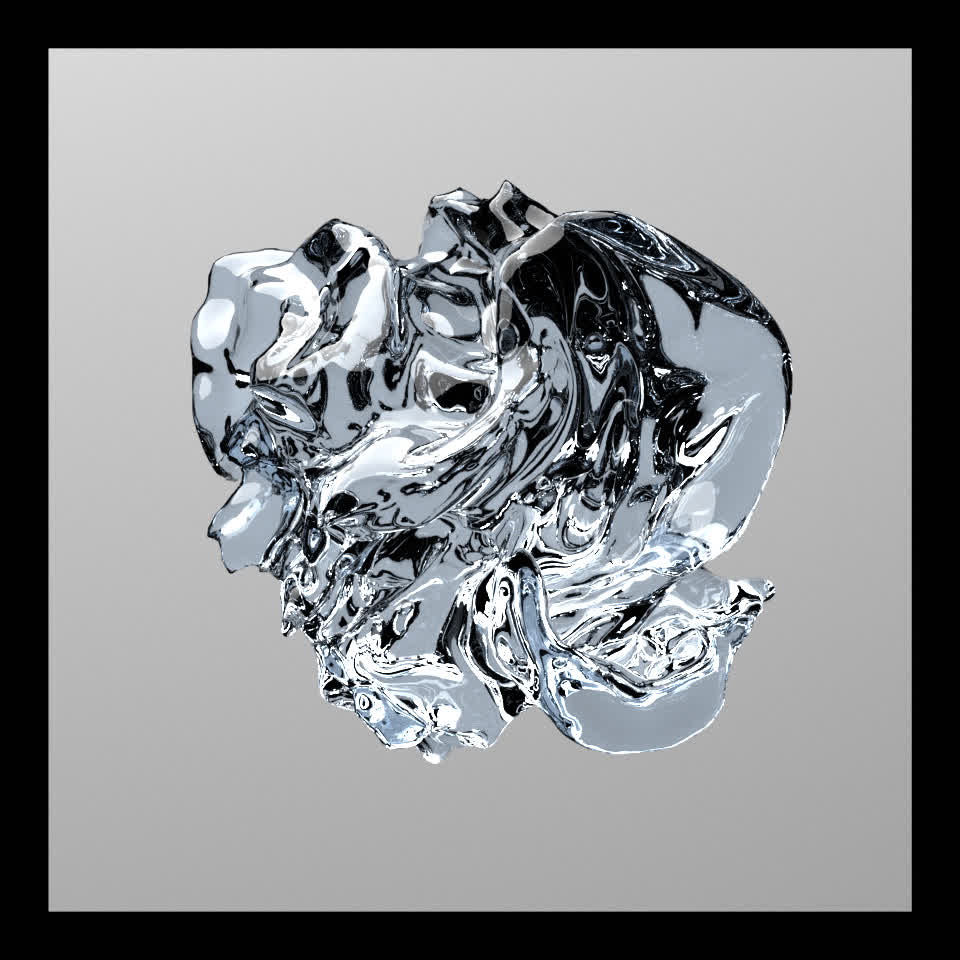} };
            \end{tikzpicture}\\ [1pt]
            $\sqrt{I_1/I_3}$ & $0.946$ & 0.633 & 0.905 \\[5pt]
            $d/d_A$          &  0.999  & 0.965 & 0.701 \\
            \end{tabular} };
        \node at (0,0) {\includegraphics[width=.55\textwidth]{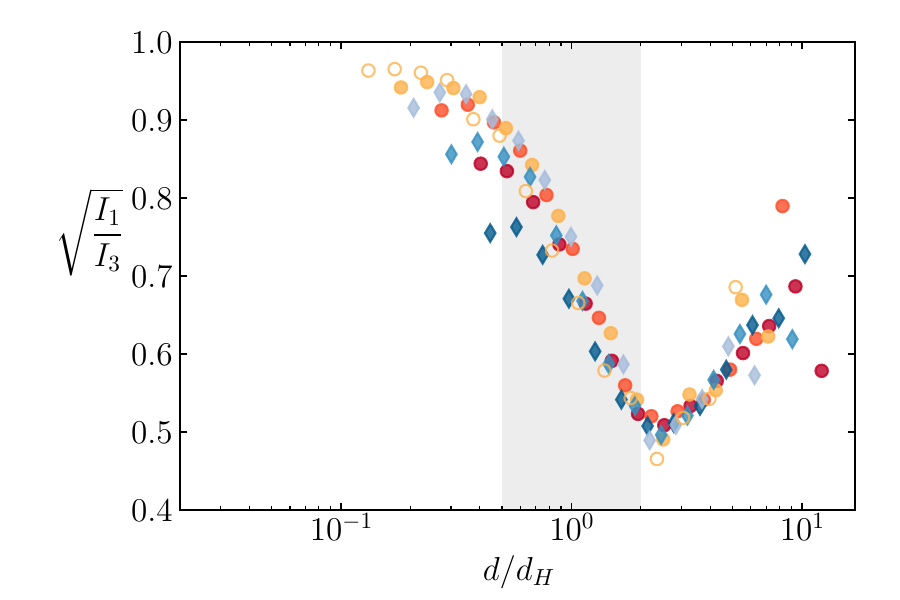}};
        \node at (7,0) {\includegraphics[width=.55\textwidth]{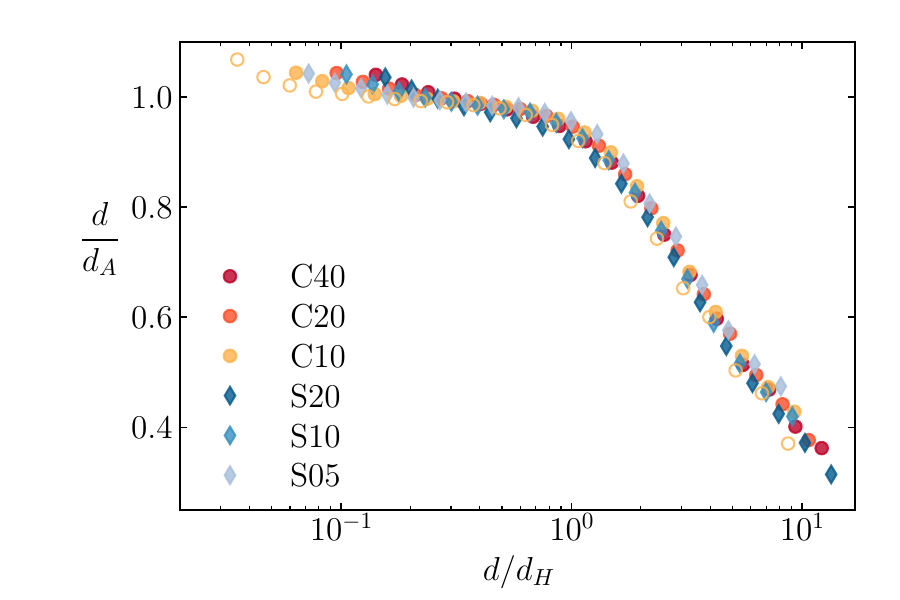}};
        \node at (-1.6,5.8){$(a)$};
        \node at (-3.0,1.8){$(b)$};
        \node at (+4.0,1.8){$(c)$};
  \end{tikzpicture}
  }
  \caption{Mean droplet deformation at each equivalent diameter $d$ normalised by the Kolmogorov-Hinze scale $d_H$. We plot two measures of deformation; $(b)$ the aspect ratio of the droplets $\sqrt{I_1/I_3}$, and $(c)$ the sphericity $d/d_A$. On panel $(b)$, we shade the region $0.5<d/d_H<2$ where the aspect ratio is observed to decrease with equivalent diameter. \ours{In both panels $(b)$ and $(c)$ we report data from the simulation performed on a fine grid, $N_f=1000$, with empty markers.}
  Panel $(a)$ above shows the values of deformation for (from left to right): a spheroid, an ellipsoid and a bulgy droplet.}
\label{fig:deformation}
\end{figure}

Next, we look at the shape of the droplets, which provides an indication of the competition between surface tension and turbulence: a droplet in a quiescent fluid takes a spherical shape and any deviation from this shape is to be attributed to the flow. We evaluate the shape of the droplets using two dimensionless parameters: the aspect ratio $\sqrt{I_1/I_3}$ and the sphericity $d/d_A$. 

The aspect ratio is computed as the ratio between the smallest and the largest eigenvalues of the moment of inertia tensor, respectively $I_1$ and $I_3$, with $I_1 \le I_2 \le I_3$. The moment of inertia tensor is computed as 
\begin{equation}
    \mathbf{I}\equiv \int_V \rho (||\mathbf{r}||^2 \mathcal{I}-\mathbf{r}\otimes \mathbf{r}) dV,
\end{equation}
where $\mathbf{r}$ is the vector from the droplet centre of mass to a point inside the droplet, and $V$ is the volume of the considered droplet. The aspect ratio uses a similar definition of the bubble deformation parameter defined in \citet{bunner2003effect}, although in their original work, the bubble deformation was defined as the square root of the largest over the smallest eigenvalue and a slightly different formulation of the moment of inertia tensor was used. Here, we choose to define the aspect ratio as the inverse of the bubble deformation found in \citet{bunner2003effect}, such that the values of the aspect ratio are bounded between 0 (e.g., infinitely long and thin filament or sheet) and 1 (e.g., sphere or cube). \ours{To exclude grid resolution effects when calculating aspect ratios, we consider only droplets with a volume of more than 100 grid cells.}

The sphericity is defined as the ratio of the volume-equivalent diameter over the surface-equivalent diameter; $d/d_A$. We use a different definition of sphericity from that found in the literature \citep{wadell_volume_1935-1}, such that it is bounded between 0 and 1. The surface-equivalent diameter is defined as the diameter of a sphere having the same surface area $A$ of the considered droplet: $d_A\equiv\sqrt{A/\pi}$. Sphericity is equal to unity for a sphere (the shape with minimal surface area for a given volume) and reduces as the droplet deforms from the spherical shape. The area $A$ of each droplet is computed by counting the number of computational cells crossed by the interface and projecting a face of the computational cell onto the local normal $\mathbf{n}$ to the interface.

We choose these two parameters as they provide very different information, as illustrated in figure~\ref{fig:deformation}(a), where we compute the aspect ratio and sphericity for three sample droplets. The aspect ratio is very sensitive to droplet-scale deformations, for instance, stretching along one of the axes, but is relatively unchanged by small-scale perturbations at the interface. Conversely, the sphericity is less sensitive to large, droplet-scale deformations but is very effective in revealing small-scale perturbations of the interface. Qualitatively, the aspect ratio estimates the shape of a box that bounds the droplet. Meanwhile, the sphericity measures the total area of the droplet interface. This is reflected in figure~\ref{fig:deformation}(a): taking the spheroidal droplet as a reference ($\sqrt{I_1/I_3}\approx 1$ and $d/d_A\approx1$), the ellipsoidal droplet shows a negligible change in the sphericity value and a $\sim30\%$ reduction in the aspect ratio. The bulgy droplet, being rather compact, has a value of the aspect ratio relatively close to that of the spheroidal droplet (about 10\% difference) but a much smaller value of sphericity ($\sim30\%$ smaller). 

We report the aspect ratio $\sqrt{I_1/I_3}$ in figure~\ref{fig:deformation}(b) as a function of the droplet size normalized by the Kolmogorov-Hinze scale. We observe that making the droplet size dimensionless using the Kolmogorov-Hinze scale yields a collapse of the aspect ratio for all cases onto a single curve. 
This suggests that even in the presence of surfactants, \citeauthor{Hinze1955}'s (1955) assumptions hold, and droplet deformation is a universal function of $d/d_H$. We observe three different regimes for the aspect ratio, which can be distinguished based on the value of $d/d_H$. 
The aspect ratio of small droplets, up to approximatively $d=0.5d_H$, is roughly constant and close to unity, indicating that the droplets are spherical or only slightly elongated (i.e., compact shape). At this point, no information can yet be inferred on local, small-scale deformations of the interface. Droplets smaller than the Kolmogorov-Hinze scale are characterised by surface tension forces dominating over turbulent fluctuations.
The second regime, observed for $0.5d_H \lesssim d \lesssim 2 d_H$, is characterised by a sharp reduction in the value of the aspect ratio (down to $\sqrt{I_1/I_3}\approx 0.5$, a similar value of an elongated ellipsoid): droplets become more elongated with an overall deformation that increases with their size. This result is coherent with the definition of the Kolmogorov-Hinze scale as the size of the largest (on average) non-breaking droplets: at about the Kolmogorov-Hinze scale, droplets start to deform significantly. Statistics on the sphericity will provide further information on the small-scale deformation within this regime and will be discussed in the following paragraph.
Lastly, in larger droplets, $d>2d_H$, there is a recovery of the droplet aspect ratio: droplets partially recover from their deformed state, with values of the aspect ratio nearing $\sqrt{I_1/I_3}\approx 0.7$. This result indicates that droplets become less deformed overall, and some rotational symmetry is restored. At first, this may be a counter-intuitive result, but sphericity will help to explain this interesting behaviour. 
\ours{We include also data from the refined grid simulation (empty markers) and we report that it closely follows the standard grid data.}

We report the sphericity for all cases in figure~\ref{fig:deformation}(c). The droplet size is made non-dimensional using the Kolmogorov-Hinze scale for each case. We observe that data for all cases collapse on a single curve when scaled by the Kolmogorov-Hinze scale, further confirming the validity of the Kolmogorov-Hinze scale as a fundamental length scale. We observe two very distinct regimes, separated by the Kolmogorov-Hinze scale. At scales smaller than the Kolmogorov-Hinze scale, droplets have a close-to-unity and slightly decreasing sphericity, which sharply decreases above the Kolmogorov-Hinze scale. Note that, for the smallest droplets, we have values of sphericity larger than one; we would like to remark that these values are not admissible and are due to inaccuracies in the computation of interface normals and area when the droplets are only a few grid cells in volume.
The sphericity for droplets smaller than the Kolmogorov-Hinze scale is close to unity and decreases for increasing droplet sizes; this information, coupled with the results from the droplet aspect ratio, indicates that droplets much smaller than Kolmogorov-Hinze scale have a spheroidal shape with limited elongation and almost no small-scale perturbations of the interface. Kolmogorov-Hinze-scale-sized droplets (but still smaller than the Kolmogorov-Hinze scale) show a substantial reduction in the aspect ratio and only a minor decrease in the sphericity: the shape of these droplets is similar to an ellipsoid, as the droplet is stretched (low aspect ratio) but the sphericity is still close to unity (indicating the absence of relevant perturbations of the interface).
Conversely, droplets slightly larger than the Kolmogorov-Hinze scale show reduced aspect ratio and sphericity: these droplets are not only strongly elongated (low aspect ratio), but small-scale perturbations of the interfaces (small humps and dimples) start forming (low sphericity). The trend in sphericity is kept also for droplets much larger than the Kolmogorov-Hinze scale; these droplets show a recovery of the aspect ratio, indicating either the formation of bulgy droplets (see figure~\ref{fig:deformation}a) or of convoluted filaments. Both of these shapes are coherent with the two deformation parameters we investigated. i.e., relatively low aspect ratio and low sphericity.
\ours{Data from the refined grid simulation (empty markers) collapses well onto the standard grid data, confirming grid-independence of the results. Due to the higher grid resolution, the fine grid case is able to capture smaller droplets compared to the standard case, however the general trend is kept.}

\begin{figure}
  \centerline{
   \includegraphics[width=0.55\textwidth]{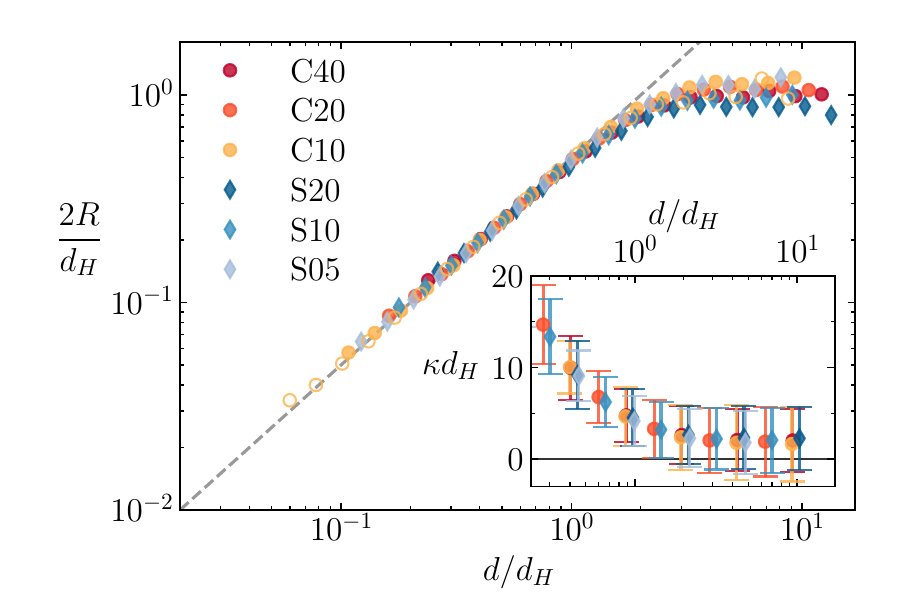}
  }
  \caption{The mean radius of curvature $R$ of the surfaces of droplets binned by their diameter $d$. The dashed line shows $R=d/4$, the radius of curvature of a sphere with diameter $d$; \ours{data from the simulation performed on a fine grid, $N_f=1000$, is reported with empty markers}. Inset: the same values, plotted in terms of curvature $\kappa=1/R$. Error bars show the standard deviation of curvature on the droplets.}
\label{fig:kurv}
\end{figure}

To distinguish among the two possible shapes, bulgy droplets versus convoluted filaments, we compute the average radius of curvature $R$ at the interface of each droplet, reported in figure~\ref{fig:kurv}.
To compute the radius of curvature $R$, we first compute the mean curvature $\kappa$ using the divergence of the normal $n$ to the interface; 
\begin{equation}
    \label{eq:curv}
    \kappa=\nabla\cdot\mathbf{n}.
\end{equation}
We average the curvature $\kappa$ over the droplet interface, and define the radius of curvature as the inverse of the mean curvature, $R\equiv 1/\kappa$. Note that, in equation~\ref{eq:curv}, we do not use a minus sign in the definition of the curvature, as we have an outward-pointing normal $\mathbf{n}$ and we choose to assign positive values of curvature to convex surfaces, i.e., the curvature is positive if the surface curves away from the normal. Also, the normal is ill-defined for small droplets, so we only calculate the mean curvature of droplets with a volume greater than 10 computational cells. For a surface in three dimensions, such as our interface, the mean curvature is equal to the sum of the two principal curvatures, $\kappa=\kappa_1+\kappa_2$. For a sphere, both principal curvatures are equal to the inverse of the sphere's radius; hence the mean curvature is twice this value and the radius of curvature is $R=d/4$. Figure~\ref{fig:kurv} shows that droplets smaller than the Kolmogorov-Hinze scale follow this scaling (gray dashed line). This result further confirms the rather regular shape (spheroid- or ellipsoid-like) of the small droplets. Above the Kolmogorov-Hinze scale instead, we observe a departure from $d/4$: the radius of curvature becomes constant, approximately equal to half the Kolmogorov-Hinze scale, and independent of drop size. 
\ours{The good agreement obtained with data from the fine grid simulation (empty markers) indicates that the grid resolution is more than sufficient to capture the local shape of the interface.}
 
To understand this behaviour, we consider a cylinder of radius $d_H/2$. On the curved surface of the cylinder, the two principal curvatures are $\kappa_1=0$ and $\kappa_2=2/d_H$. Neglecting the two flat ends, the mean curvature of the cylinder is thus $\kappa=2/d_H$, i.e. the radius of curvature of the cylinder is equal to half the Kolmogorov-Hinze scale, and is independent of its length. This result, combined with the information obtained from the deformation parameters in figure~\ref{fig:deformation}, shows that, above Kolmogorov-Hinze scale droplets take the shape of filaments with a diameter equal to the Kolmogorov-Hinze scale.

The inset of figure~\ref{fig:kurv} shows the mean curvature $\kappa$ of the droplets as a function of their equivalent diameter $d$. Error bars indicate the standard deviation of the curvature, which was calculated at the interface of each droplet and averaged over the ensemble of droplets of similar size $d$. The standard deviation of the curvature of an interface is a measure of its corrugation, and can be used to quantify Plateau–Rayleigh instabilities which lead to droplet breakup \citep{rayleigh_instability_1878,villermaux_single-drop_2009,kooij_what_2018}. From the inset of figure~\ref{fig:kurv}, we see that droplets of all sizes show corrugation, and the standard deviation of $\kappa$ is comparable to its average value. Above the Kolmogorov-Hinze scale, we report an increased probability of negative values of the mean curvature, indicative of dimples and saddle points in the surface of the droplet, which, for example, can be due to impinging jets or large Plateau-Rayleigh instabilities on the interface.

\begin{figure}
  \centerline{
   \includegraphics[width=0.55\textwidth]{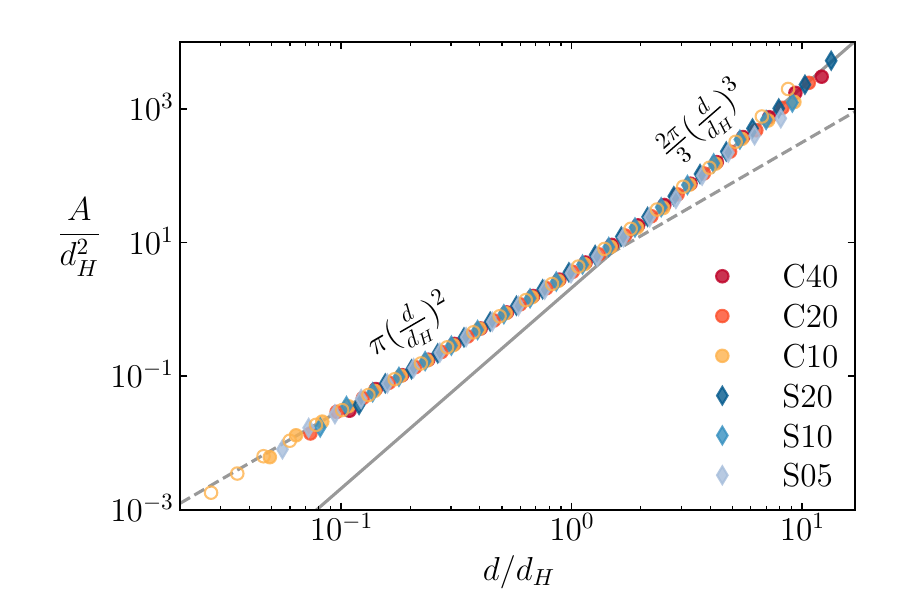}
  }
  \caption{Dependence of droplets' surface area $A$ on their equivalent diameter $d$. The dashed line shows the surface area of a sphere with diameter $d$, and the solid line shows the surface area of a filament with diameter $d_H$. \ours{Data from the simulation performed on a fine grid, $N_f=1000$, is reported with empty markers}}
\label{fig:area}
\end{figure}

To test our hypotheses on the droplet shapes above and below the Kolmogorov-Hinze scale, we consider their surface area. 
Figure~\ref{fig:area} shows the mean surface area of the droplets at each equivalent diameter $d$. As we previously noted, the droplets smaller than the Kolmogorov-Hinze scale are almost spherical in shape, and hence we see that their surface area grows quadratically with their characteristic size. 
Above the Kolmogorov-Hinze scale, the droplet surface areas instead grow as the cube of their sizes. 
As a first approximation, we can think of these droplets as long cylinders with the same diameter $d_H$ and different lengths $l$. Ignoring the two end faces, the surface area of such a cylinder is $A_f=\pi d_H l$ and the volume is $V_f = \pi d_H^2 l/4$. Substituting the volume for the equivalent diameter (equation~\ref{eq:diam}) we get an expression for the length;
\begin{equation}
\label{eq:filamentVol}
l = \frac{2d^3}{3d_H^2}.
\end{equation}
Plugging this into the formula for the cylinder surface area, we obtain $A_f=2\pi d^3/(3d_H)$, showing that a filament with variable length $l$ and constant diameter $d_H$ has an interfacial surface area that grows as the cube of its volume-equivalent diameter $d$. The interfacial surface area of the droplets above the Kolmogorov-Hinze scale closely follows $A_f$, as reported in figure~\ref{fig:area}. 
\ours{The two scalings are also confirmed by simulation on a more refined grid (empty markers).}

\begin{figure}
  \centerline{  
  \begin{tikzpicture}[x=6cm, y=6cm, font=\footnotesize]
      \node[anchor=south west,inner sep=0] (image) at (0,0) {\adjincludegraphics[width=6cm,trim={{.05\width} {.05\width} {.05\width} {.05\width}},clip]{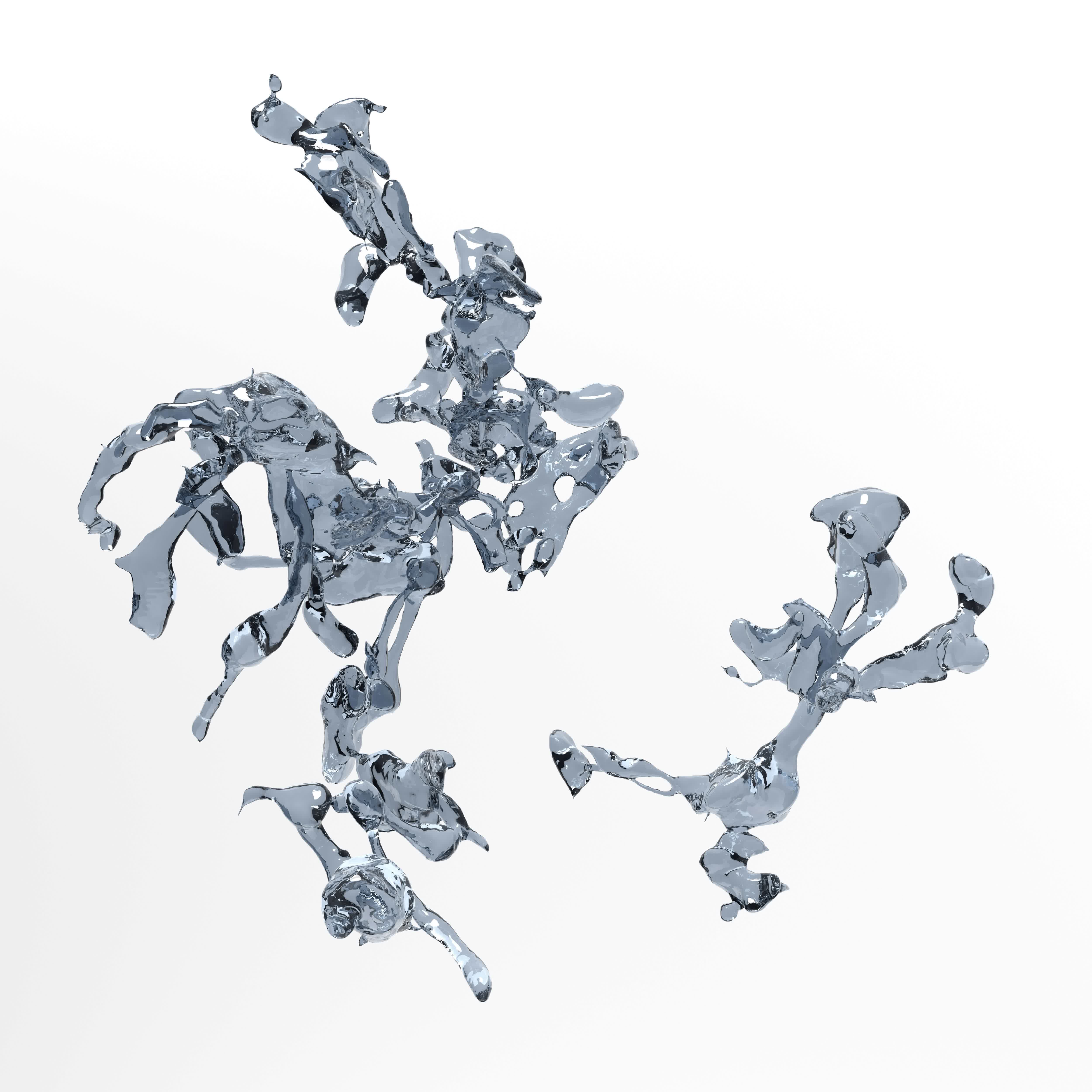}};
      \fill (image.south east) ++(-12mm,1mm) rectangle ++(.0232/.9*482/432*781/581, .5mm) 
      node[anchor=west, shift={(0mm,0mm)}] {$d_H$};
  \end{tikzpicture}    
 }
  \caption{Visualisation of two droplets extracted from case S20. The left droplet has 56 handles and six voids. The right droplet has two handles and zero voids. The length of the black bar is the Kolmogorov-Hinze scale.}
\label{fig:handle56}
\end{figure}

\begin{figure}
  \centerline{
  \begin{tikzpicture}
    \node at (0,0) {\includegraphics[width=.55\textwidth]{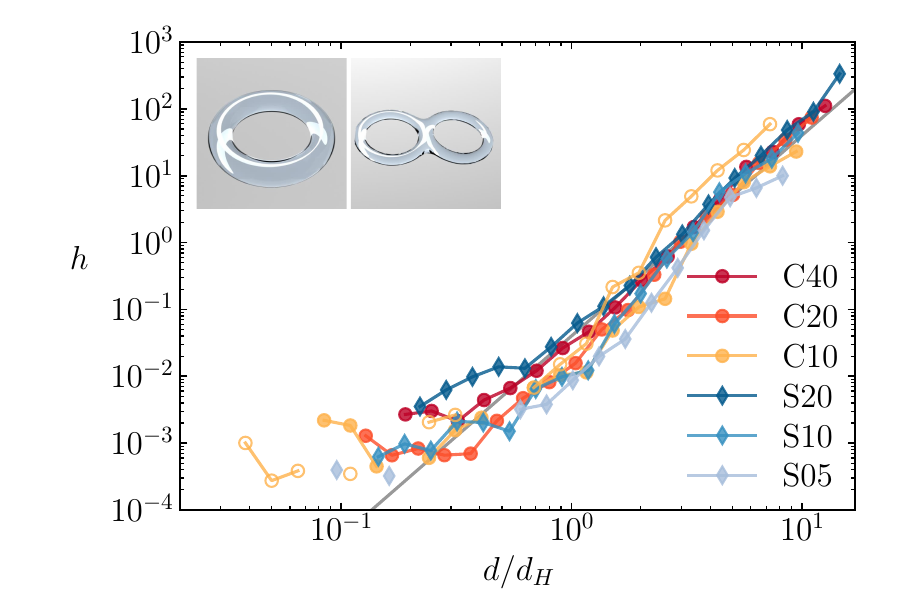}};
    \node at (7,0) {\includegraphics[width=.55\textwidth]{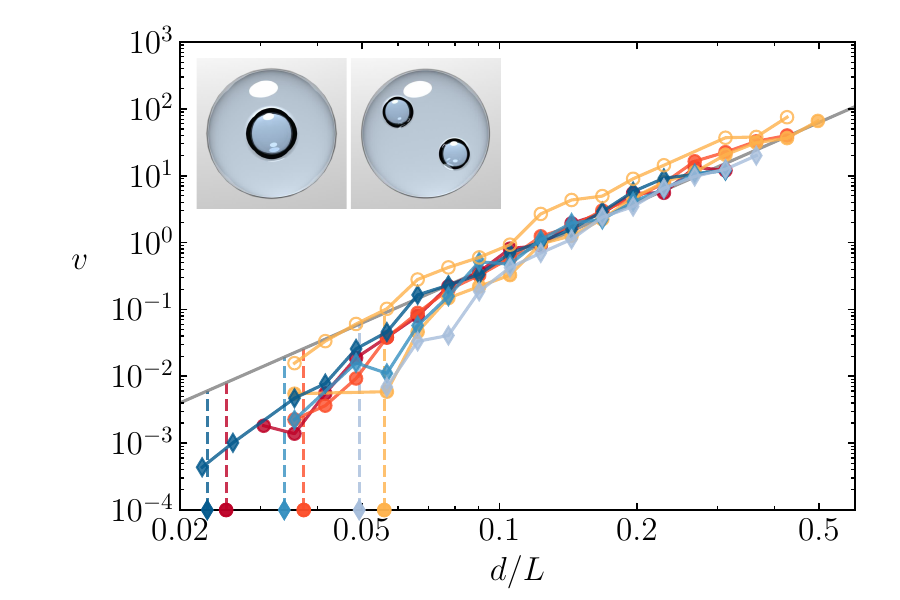}};
    \node at (-3.0,1.8){$(a)$};
    \node at (+4.0,1.8){$(b)$};
  \end{tikzpicture}
  }
  \caption{$(a)$ Number of handles $h$ per droplet. The grey line shows the fit $h=0.04(d/d_H)^3$. Renders show example droplets with one and two handles ($\chi=0$ and $\chi=-2$, respectively). $(b)$ Number of voids $v$ per droplet. The grey line shows the fit $v=500(d/L)^3$. We mark the x-axis to show the Kolmogorov-Hinze scale of each case. \ours{We report data from the simulation performed on a fine grid, $N_f=1000$, using empty markers.} Renders show example droplets with one and two voids ($\chi=4$ and $\chi=6$, respectively). The renders are obtained from droplets we artificially generated for the sole purpose of demonstrating handles and voids.}
\label{fig:genus}
\end{figure}

From our observations of curvature and surface area, it appears that droplets above the Kolmogorov-Hinze scale are filamentous in shape. However, their aspect ratio shows the filaments cannot be straight, since this would produce a monotonic reduction in $\sqrt{I_1/I_3}$ with $d$, which we do not see in figure~\ref{fig:deformation}a. Hence a natural question is: \textit{do the filaments form loops, or are they simply connected?} Figure~\ref{fig:handle56} shows two droplets from case S20, we see that both droplets are made up of convoluted filaments, and in many places, the filaments do in fact, form loops. 
\reva{These complex shapes are typical of large and deformable viscous droplets in both liquid-liquid mixtures \citep{Andersson2006,Collins1970,li2017size,xue2019formation} and gas-liquid mixtures \citep{blenkinsopp2010bubble,dumouchel2014laser,cheron_analysis_2022,kooij_what_2018,Hinze1955,jain2015secondary,qi_fragmentation_2022, villermaux_fragmentation_2007,kant_bag-mediated_2023}. Small droplets (smaller than the Kolmogorov-Hinze scale) have instead a more spherical shape, as indicated in figure~\ref{fig:deformation}.}
To answer the question quantitatively, we measure the topology of the interface of each droplet using its Euler characteristic $\chi$, which obeys the formula 
\begin{equation}
1-\chi/2=h-v, 
\label{eq:handles}
\end{equation}
where $h$ is the number of handles and $v$ is the number of voids in the drop (see section~\ref{sec:Euler} in the appendix for a derivation of this equation). In the insets of figure~\ref{fig:genus}a, we show renders of example droplets with one and two handles. Analogously to the handle on a teacup, a handle is a loop of the dispersed phase which extends through the carrier phase. The renders in the insets of figure~\ref{fig:genus}b show example droplets with one and two voids. A void is a region of the carrier phase entirely enclosed by the dispersed phase. Note that similarly to droplet breakup and coalescence, a change in the number of handles or voids necessitates a merging or splitting of interfaces. Using a method similar to \citeauthor{mendoza_evolution_2006}'s characterisation of dendritic metal samples, we measure the Euler characteristic using simplicial homology, that is, by dividing the interface into simple polygons and counting the number of nodes $n$, edges $e$, and faces $f$ on the interface of each droplet. The Euler characteristic of the interface is then given by the Poincar\'e formula \citep[p.26]{massey_basic_1997},
\begin{equation}
\chi=n-e+f.
\label{eq:euler}
\end{equation}
Our volume of fluid field $\phi$ is defined on a cubic grid, and we can define the interface as the boundaries between cells where $\phi-0.5$ changes sign. Hence, our interface is already divided into square faces, and $\chi$ can be calculated by counting these faces and their edges and nodes. 
We also count the number of voids $v$ on each droplet. Voids are counted by rerunning the flood fill algorithm, looking for contiguous cells where $\phi<0.5$; to do so we use a stack-based 26-way flood-fill, which is an extension to three dimensions of the two-dimensional 8-way flood fill \citep{newman1979principles}. Knowing $\chi$ and $v$ for each drop, we can use equation~\ref{eq:handles} to obtain the number of handles $h$.

Figure~\ref{fig:genus}a shows how the mean number of handles per droplet depends on the droplet size. We see that the largest droplets are very self-connected, having on the order of $10^2$ handles. Furthermore, the number of handles in each case collapses to a single line when the equivalent diameter is normalised by the Kolmogorov-Hinze scale $d_H$. This universality occurs because surface tension is constantly acting to destroy the handles, so the higher the surface tension, the shorter the lifetime of the handle. The fit $h=0.04(d/d_H)^3$ gives the number of handles as a function of the droplet volume. 
Using our earlier result that the large droplets are filaments with lengths given by equation~\ref{eq:filamentVol}, we can convert this expression into the number of handles per unit length of the filament: $h/l=0.06/d_H$.
Figure~\ref{fig:genus}b shows us, on the other hand, that the number of voids is independent of surface tension, which is reasonable as a void inside a droplet experiences no net surface tension force. The number of voids simply scales with the volume of the droplet. Again we can obtain a fit, $v=500(d/L)^3$; substituting the equivalent diameter for the droplet volume $V$ (equation~\ref{eq:diam}) we find there are roughly $v/V\approx950/L^3$ voids per unit droplet volume in all cases. We suspect the void concentration depends on the droplet coalescence rate and turbulence intensity ($Re_\lambda$). However, we leave a proper investigation of this dependence to future works.

\ours{Data from the fine grid simulation is reported with empty markers for both handles and voids, and a good agreement with data computed on the standard grid is observed. 
When counting the number of voids in the fine grid case, we excluded voids with a volume smaller than eight computational cells,
(i.e. one computational cell on the standard grid). This method allows for a fair comparison of the number of voids in the fine grid and the standard grid cases.
}

\subsection{Flow statistics} 

\begin{figure}
  \centerline{
  \begin{tikzpicture}
    \node at (0,0) {\includegraphics[width=.48\textwidth]{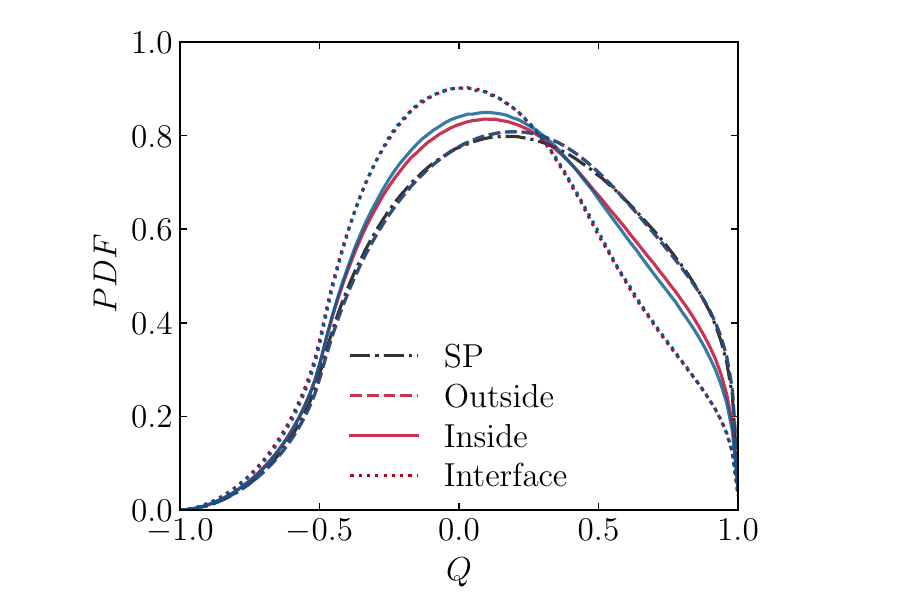}};
    \node at (4.4,0) {\includegraphics[width=.48\textwidth]{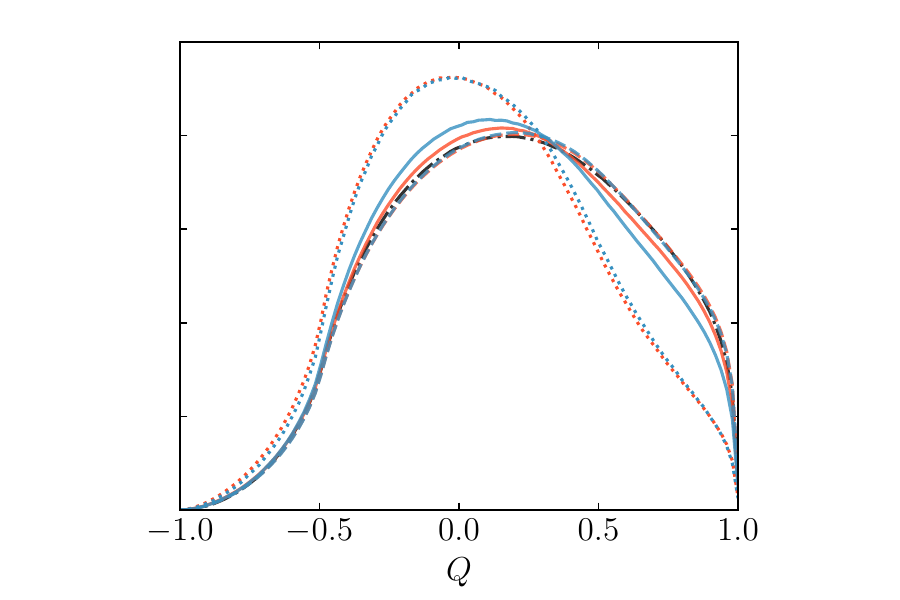}};
    \node at (8.8,0) {\includegraphics[width=.48\textwidth]{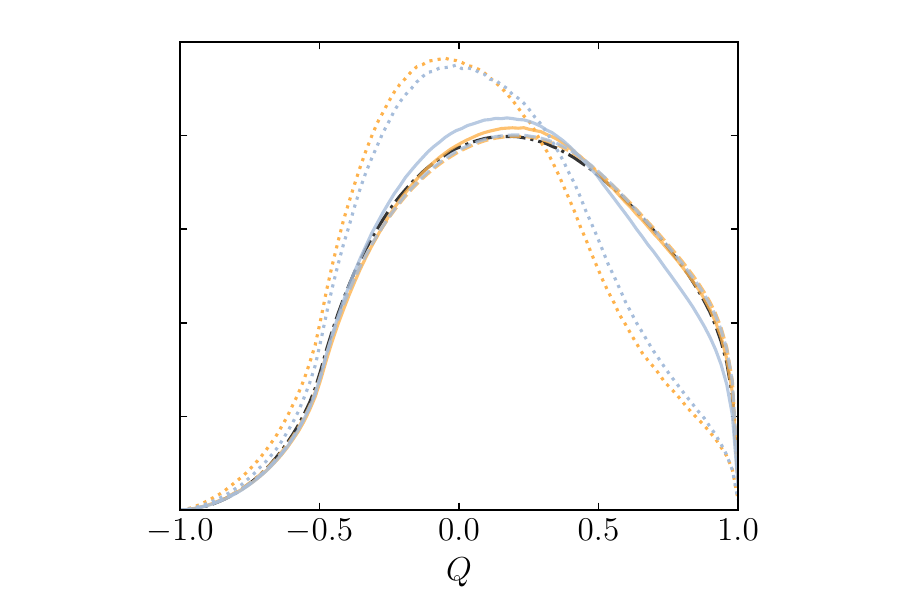}};
    \node at (-1.55,1.5){$(a)$};
    \node at (+2.85,1.5){$(b)$};
    \node at (+7.25,1.5){$(c)$};
  \end{tikzpicture}
  }
  \caption{Flow topology parameter sampled in different regions of the domain: inside the droplets (solid line), in the carrier phase (dashed line) and at the interface (dotted line). The single-phase case is reported for reference (dash-dotted line). Each panel refers to an approximate value of the effective Weber: $(a)$ $We_e\approx 10$ -- panel (cases C10 and S05), $(b)$ $We_e\approx 20$ -- panel (cases C20 and S10) and $(c)$ $We_e\approx 40$ -- panel (cases C40 and S20). }
\label{fig:topology}
\end{figure}
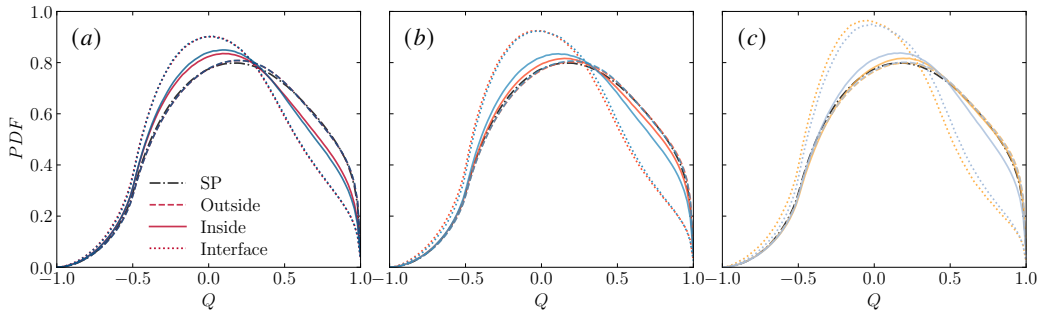

We characterize the effects induced by the presence of clean and surfactant-laden interfaces on the local flow statistics using the flow topology parameter \citep{perry1987description}, which compares the local flow to three different base flows: purely rotational, pure shear and purely extensional flow. The flow topology parameter $Q$ is a combination of the rate-of-\ours{strain} tensor $S\equiv(\nabla \mathbf{u}+\nabla \mathbf{u}^T)/2$ and of the rate-of-rotation tensor $\Omega\equiv(\nabla \mathbf{u}-\nabla \mathbf{u}^T)/2$, where $\nabla\mathbf{u}$ is the velocity gradient tensor. The quantities $S^2$ and $\Omega^2$ are defined as $S^2=S:S$ and $\Omega^2=\Omega:\Omega$, where $:$ identifies the dyadic (double-dot) product. For $S^2=0$ and $\Omega^2 \ne 0$ we have a purely rotational flow, whereas for $S^2\ne0$ and $\Omega^2 = 0$ we have a purely extensional flow; pure shear is a combination of these two cases, occurring for $S^2=\Omega^2$. 
\begin{equation}
    Q=\frac{S^2-\Omega^2}{S^2+\Omega^2}= 
    \begin{cases}
        -1 & \text{purely rotational,}\\
         0 & \text{pure shear,}\\
        +1 & \text{purely extensional.}
    \end{cases}
\end{equation}

We compare clean and surfactant-laden cases at similar effective Weber numbers in figure~\ref{fig:topology}, together with the single-phase case for reference. The flow topology parameter is computed in three distinct regions: inside the droplets ($\phi>0.5$), in the carrier phase ($\phi<0.5$) and at the interface. This way, we can separate the contribution from the different regions of the flow \citep{dodd2019small,RostiDB_2019,soligo2020effect}, and investigate the effect of Marangoni stresses at the interface for the surfactant-laden cases and of flow confinement on the flowing condition inside the droplets and at the interface. 

We first consider the flow topology parameter in the carrier phase. The relatively low volume fraction of the dispersed phase reduces the overall impact of the presence of the interface on the outer flow. The flow topology parameter for all clean and surfactant-laden cases well collapses onto the single phase line, indicating indeed that the presence of a deformable interface and of Marangoni stresses does not introduce any significant modification of the outer flow at the relatively low volume fraction considered. A similar result was reported by \citet{RostiDB_2019} for clean droplets and for volume fractions of the dispersed phase up to 30\%.
In general, the flow topology shows a predominance of a combination of pure shear and extensional flow, with a limited rotational contribution.

The flow topology computed inside the droplets highlights the effect of flow confinement. Surface tension has competing effects: while on the one hand, a large value of surface tension is more effective in decoupling internal and external flow, a low value of surface tension produces many small-size droplets that increase the flow-confinement effect. We observe a reduction in extensional flow and an increase in pure shear for increasing values of the Weber number, which has been attributed to the confinement effect of small droplets \citep{soligo2020effect}. We do not report changes in the rotational component, which can be attributed to the lack of large-scale coalescence events \citep{RostiDB_2019}. When comparing cases at similar effective Weber numbers, we observe that the addition of surfactant has a similar effect to a reduction in the surface tension value: a reduction in the extensional flow and an increase in pure shear. The difference in the flow topology between clean and surfactant-free droplets reduces as the effective Weber number increases.

It is interesting to note that, at the interface, the addition of surfactant instead has an opposite effect: the presence of surfactant, especially at low values of the effective Weber number, leads to an increase of extensional flow and a decrease in rotational flow -- pure shear is unchanged. This difference results from the action of Marangoni stresses, and is indeed more apparent at low values of the effective Weber number, i.e. at high values of surface tension. We observe a shift towards extensional flow for increasing values of the effective Weber number, confirming previous findings \citep{soligo2020effect}.

\begin{figure}
  \centerline{
  \begin{tikzpicture}
    \node at (0,0) {\includegraphics[width=.55\textwidth]{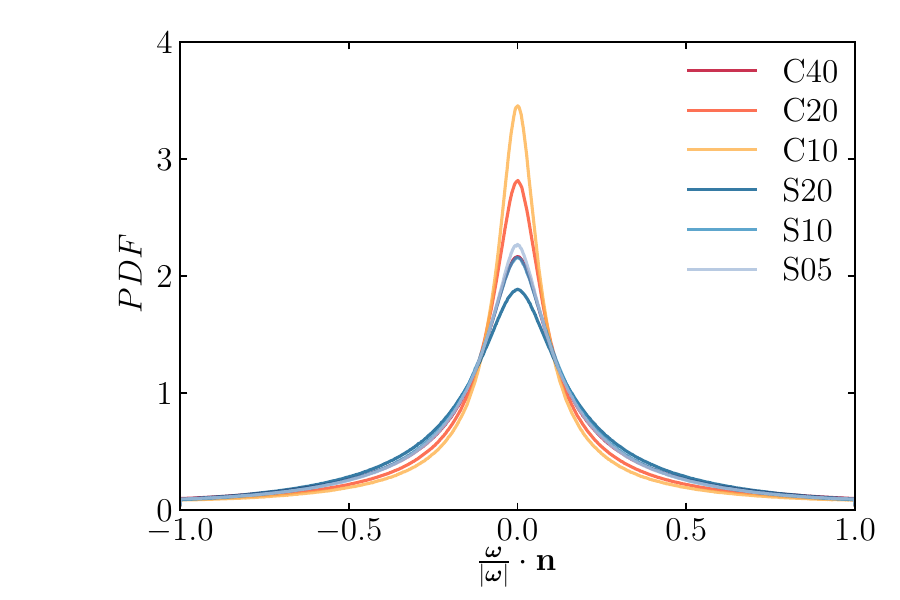}};
    \node at (7,0) {\includegraphics[width=.55\textwidth]{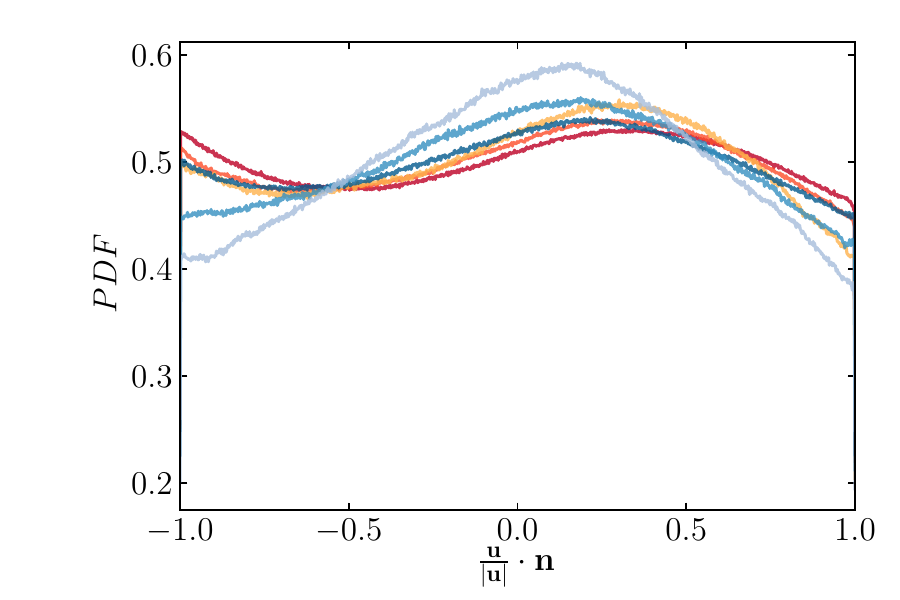}};
    \node at (-1,1) {\includegraphics[width=.13\textwidth]{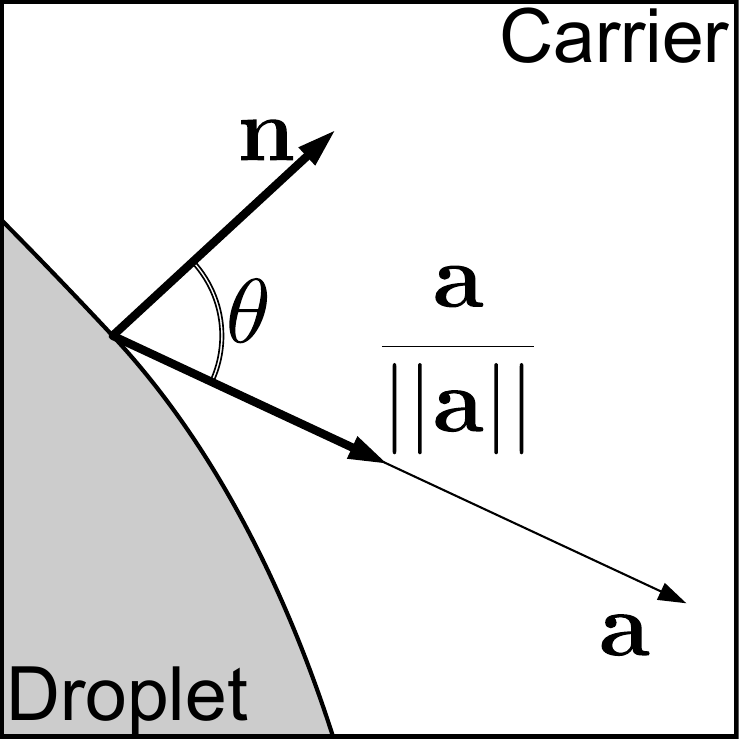}};
    \node at (-3.0,1.8){$(a)$};
    \node at (+4.0,1.8){$(b)$};
  \end{tikzpicture}
  }
  \caption{Alignment between unit-length normal to the interface $\mathbf{n}$ and: $(a)$ vorticity at the interface, $(b)$ velocity at the interface. The inset in panel $(a)$ shows how the alignment has been computed: for a generic vector $\mathbf{a}$ we define the alignment as $\mathbf{n}\cdot \mathbf{a}/||\mathbf{a}||$, which is equal to the cosine of the angle $\theta$ in between the two vectors. 
  }
\label{fig:alignment}
\end{figure}

We now focus on flow statistics at the interface to better understand the local surface tension effects induced by clean and surfactant-laden interfaces. 
We compute the alignment of vorticity at the interface of the droplets: to quantify the direction of vorticity, we use the cosine of the angle $\theta$, see the sketch in figure~\ref{fig:alignment}a. This quantity is found by taking the scalar product between the interface normal $\mathbf{n}$ (outward-pointing, unit-length normal) and the unit-length vorticity vector, $\boldsymbol{\omega}/||\boldsymbol\omega||$. The probability density function of the vorticity-interface alignment is reported in figure~\ref{fig:alignment}a for all the cases we simulated. Similarly to what was found by \citet{mukherjee2019droplet}, the vorticity is mostly orthogonal to the interface normal. At very large values of the surface tension, i.e. $We\to 0$, the interface between the droplet and the carrier phases acts similarly to a slip wall: the high surface tension makes the interface close to undeformable but imposes no condition on the tangential component of the flow. 
We thus expect that at high values of surface tension the vorticity vector is tangential to the interface, i.e. $\mathbf{n}\cdot\boldsymbol\omega/||\boldsymbol\omega||=0$. As the interface becomes more deformable (corresponding to a higher Weber number), this condition is relaxed and the PDF widens. We observe that the PDF remains symmetric for all cases; this is an expected result as positive and negative values of the alignment correspond to flow structures at the interface rotating in the anti-clockwise and clockwise directions respectively, and the two are equally probable. 
The alignment of vorticity at the interface also highlights the effect of surfactant, and in particular of Marangoni stresses tangential to the interface: the cases at a similar effective Weber number (namely $We_e\approx 10$ C10 and S05; $We_e\approx 20$ C20 and S10; $We_e\approx 40$ C40 and S20) show notably different distributions, suggesting that in terms of the local flow around the droplets, the effect of surfactant cannot be approximated as only an average surface tension reduction. We indeed notice a decoupling among the various cases at approximately the same effective Weber number, with cases C40 ($We_e=39.3$) and S10 ($We_e=25.8$) showing the very same distribution in vorticity alignment. The presence of Marangoni stresses promotes the formation of flows tangential to the interface, whose gradients contribute to the vorticity component normal to the interface.

The alignment of the fluid velocity at the interface further corroborates the role of Marangoni stresses in modifying the local flow velocity at the interface; figure~\ref{fig:alignment}b shows the probability density function of the scalar product between the unit-length velocity vector and the normal to the interface. The surfactant-laden cases show a more peaked distribution at $\mathbf{n}\cdot \mathbf{u}/||\mathbf{u}||\approx 0.1$, corresponding to a local fluid velocity almost tangential to the interface, with a small outward (i.e., from the droplet phase towards the carrier phase) component. This velocity alignment is clearer for the cases at low Weber number: the magnitude of Marangoni stresses directly depends on the local surface tension, hence the cases at high Weber number are characterized by weaker Marangoni stresses. Indeed, the case S20 shows similar velocity alignment to the surfactant-free cases, being the Marangoni stresses weaker compared to the other surfactant-laden cases. For the clean cases we observe two separate peaks in the distribution, one at $\mathbf{n}\cdot \mathbf{u}/||\mathbf{u}|| = -1$ and one at $\mathbf{n}\cdot \mathbf{u}/||\mathbf{u}||\approx 0.3$. The former corresponds to an inward flow perpendicular to the interface, and the latter to a fluid velocity mainly tangential to the interface, although with a larger normal component compared to the surfactant-laden cases. We attribute this reduction in the probability of having flow tangential to the interface to the absence of Marangoni stresses for the surfactant-free cases. A similar distribution is also achieved by case S20, which is characterised by weak Marangoni stresses (due to the low reference surface tension value), further highlighting the role of Marangoni stresses. 

\begin{figure}
  \centerline{
  \begin{tikzpicture}
    \node at (0,0) {\includegraphics[width=.55\textwidth]{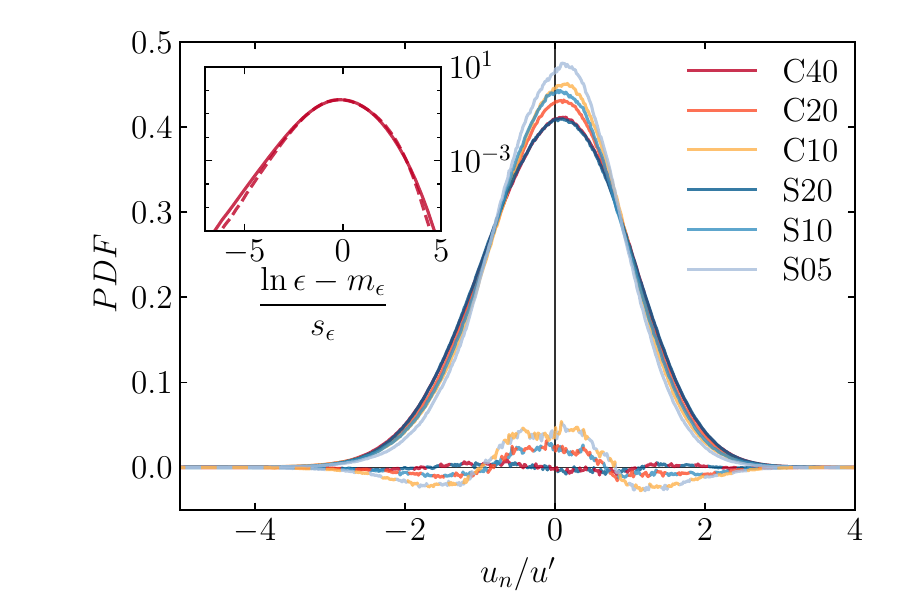}};
    \node at (7,0) {\includegraphics[width=.55\textwidth]{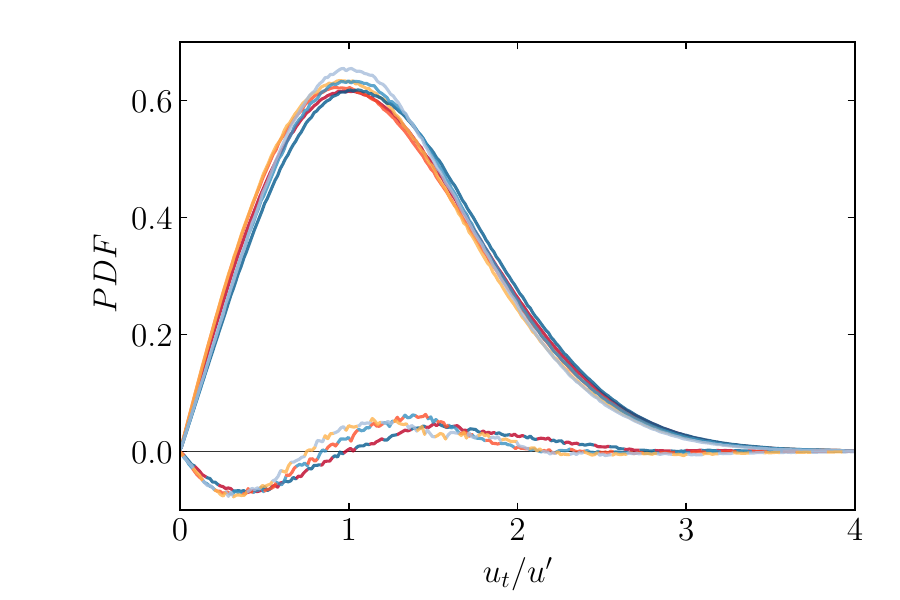}};
    \node at (9,1) {\includegraphics[width=.13\textwidth]{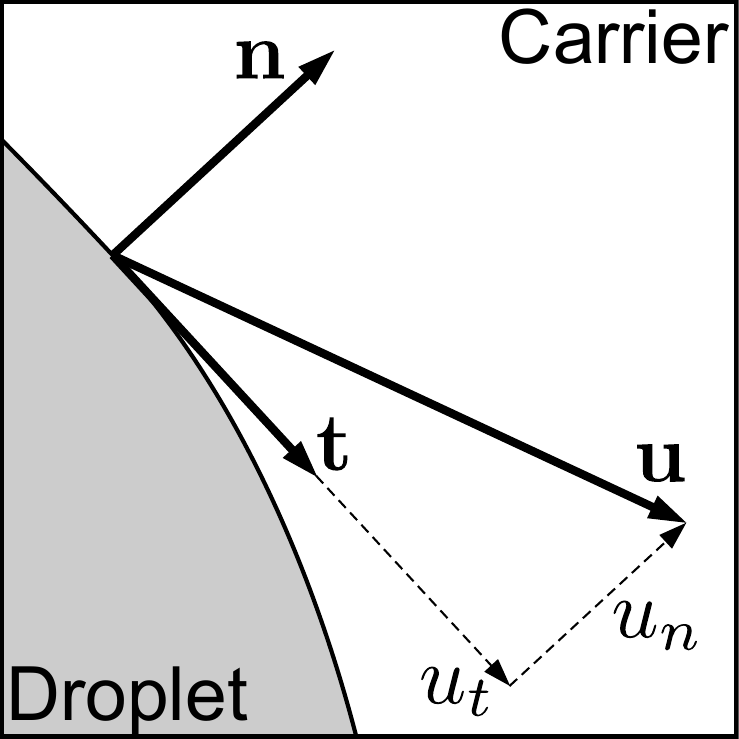}};
    \node at (-3.0,1.8){$(a)$};
    \node at (+4.0,1.8){$(b)$};
  \end{tikzpicture}
  }
  \caption{$(a)$ Normal component of the flow velocity at the interface $u_n$ and $(b)$ tangential component of the flow velocity at the interface $u_t$. Two-colour lines show the difference between the PDF of surfactant-laden and clean pairs of cases with similar $We_e$ (these data are magnified by a factor two for better reading). The inset in panel $(a)$ shows 
  histograms of the logarithm of the dissipation $\epsilon$ inside (dashed line) and outside (solid line) the droplets for case C40, where $m_\epsilon$ and $s_\epsilon$ are the mean and standard deviation of $\ln\epsilon$, respectively.
  The inset in panel $(b)$ shows the decomposition of the flow velocity along a direction normal ($\mathbf{n}$) and tangential ($\mathbf{t}$) to the interface.}
\label{fig:unut}
\end{figure}
So far we have only considered the angle between the flow velocity at the interface and the interface itself; we now proceed to analyse the magnitude of the flow velocity at the interface. 
The flow velocity is decomposed into two components, a normal component $u_n\equiv\mathbf{u}\cdot \mathbf{n}$ aligned with the outward-pointing normal to the interface $\mathbf{n}$, and a tangential component $u_t\equiv ||\mathbf{u}-u_n \mathbf{n}||$. 
The sign of the normal component is important; the interface is advected with the flow so positive $u_n$ occurs in places where the interface moves outward in the direction of the carrier phase, and negative $u_n$ occurs in places where the interface moves inward, in the direction of the dispersed phase. Due to volume conservation of the phases, $u_n$ has zero mean. On the other hand, the choice of the tangential direction in the plane is arbitrary: it is taken as the remainder of the subtraction of the normal component from the total velocity. For this reason, we have only positive values of the tangential velocity $u_t$. 

In figure~\ref{fig:unut}$a$, the probability density functions of the normal component of the velocity $u_n$ show peaks at relatively low positive $u_n$ and are negatively skewed in all cases, i.e., extreme negative values (inward fluid velocity) are more probable than extreme positive values (outward fluid velocity). To elucidate the cause of the skewed distributions, we also show histograms of the logarithm of the dissipation~$\epsilon$ for case C40. The tails of these histograms can be attributed to extreme events and hence to intermittency in the flow \citep{kaneda_morishita_2012}. We see that the flow outside the droplets gives slightly wider tails and hence has higher intermittency than that inside the droplets (also seen by \citet{crialesi-esposito_intermittency_2023} for fluid velocity differences inside and outside droplets). Therefore, we attribute the increased likelihood of extreme inward velocities at the interface to increased intermittency of the flow in the bulk phase. 
Returning to the main panel of figure~\ref{fig:unut}$a$, we note that cases with a higher $We_e$ show a wider distribution of $u_n/u'$. A higher effective Weber number implies a more deformable interface with a lesser damping effect on the normal velocity: extreme (positive and negative) events become more probable as surface tension is reduced. The difference among surfactant-laden and clean cases at similar effective Weber numbers is reported with two-colour lines (rescaled by a factor of two for improved readability). We observe that the surfactant suppresses extreme events, especially at low values of the Weber number (high surface tension), when Marangoni stresses are greatest in magnitude. A similar turbulence suppression effect was seen by \citet{shen_effect_2004} for surfactants in free shear flows, and was attributed to the elasticity of the surfactant-laden interface.

When considering the tangential component, we observe a reversal of the trend: surfactant, via Marangoni stresses, increases the probability of flow tangential to the interface. Two-colour lines show the difference between surfactant-laden and clean cases at similar $We_e$, with the data rescaled by a factor of two for ease of reading. It is clear that the presence of surfactant increases the probability of finding tangential velocities in the range $u' \lesssim u_t \lesssim 2u'$. All surfactant-laden cases have approximatively a similar value of the peak of the distribution, which is slightly larger than that of the surfactant-free cases, indicating that surfactant-laden cases have a higher probability of large values of the tangential velocity. Interestingly, as the Weber number increases, the peak shifts to slightly higher values, and the likelihood of large tangential velocity increases as well, as shown by the two-colour lines. This result suggests that while increasing the flow velocity tangential to the interface and suppressing large normal components, Marangoni stresses also have a modulating effect on the tangential component.

\section{Conclusions}
\label{sec:concl}
In this study, we perform direct numerical simulations of surfactant-laden droplets in homogeneous isotropic turbulence. The interfacial dynamics are solved using an MTHINC volume of fluid method coupled with a phase-field-based approach to simulate surfactant dynamics. By examining droplet morphology and local flow statistics, we can shed light on the interfacial characteristics and dynamics of these complex systems.

The Kolmogorov-Hinze length scale is a fundamental quantity in multiphase flows laden with clean droplets or bubbles. \reva{Our numerical results confirm that the Kolmogorov-Hinze framework can be extended to surfactant-laden droplets by using an averaged surface tension value, thus accounting for the presence of surfactant. In the configuration we adopted in this study, the role of Marangoni stresses is minor, thus the effect of surfactant can be approximated as a simple average surface tension reduction.}
We indeed observe a collapse of most statistics when plotted as a function of $d/d_H$. We compute the droplet size distribution for all clean and surfactant-laden cases and verify that: (i) the Kolmogorov-Hinze scale effectively separates the breakage- and coalescence-dominated regimes and (ii) the power-law scaling for these two regimes can be applied to surfactant-laden droplets. The combined results on the deformation of the droplets, i.e. aspect ratio and sphericity, prove that droplets smaller than the Kolmogorov-Hinze scale have a relatively compact and regular shape (spheroid- or ellipsoid-like shapes), whereas droplets larger than the Kolmogorov-Hinze scale have coiled, filamentous shape, supporting previous observations of filamentous water drops \citep{villermaux_single-drop_2009,jackiw_aerodynamic_2021}. The filamentous droplets are found to have an average diameter which is independent of the overall droplet size $d$ and is equal to the Kolmogorov-Hinze scale, further evidencing the relevance of this length scale.

The very different shapes of large and small droplets have direct implications on the total area of the interface, which is a crucial parameter in determining the overall exchange of species, momentum and energy among the carrier and the dispersed phase. We report the existence of two regimes, separated by the Kolmogorov-Hinze scale: the area of droplets smaller than the Kolmogorov-Hinze scale is proportional to the square of the characteristic size of the droplet, whereas it is proportional to the cube of the characteristic size for droplets larger than the Kolmogorov-Hinze scale. The two different scalings can be directly traced back to the shape of the droplets, spheroid-like below the Kolmogorov-Hinze scale and filamentous above the Kolmogorov-Hinze scale. The large and filamentous droplets are coiled up, as indicated by their aspect ratio. Thus we investigate their self-connectedness, using the Euler characteristic to count the number of handles and voids on each droplet. To the best of our knowledge, this is the first time the number of handles and voids on a droplet has been measured. We find that the number of handles depends directly on the size of the droplet and its surface tension, as data from all cases collapse on a single curve when normalised by the Kolmogorov-Hinze scale; we also provide a scaling for the linear density of handles, $h/l=0.06/d_H$. Conversely, the number of voids depends on the droplet size alone. Our interpretation is that the restoring action of surface tension reduces the lifetime of a handle, whereas the dynamics of a void are unaffected by surface tension. Going further, the Poincar\'e-Hopf theorem relates the Euler characteristic $\chi$ of an interface to the number of topological defects in any tangent vector fields (e.g., fluid velocity, alignment of molecules, stresses) at the interface \citep{maroudas-sacks_topological_2021}, (a well-known example of this for $\chi=2$ is the hairy ball theorem). A future investigation may count topological defects on the surface of droplets and relate the number to their ability to resist breakup.

\reva{Results from the morphology of the droplets highlight the validity of the Kolmogorov-Hinze scale for both clean and surfactant-laden flows: a rescaled value of the surface tension, accounting for the average surface tension reduction induced by the surfactant, can be effectively used to define the Kolmogorov-Hinze scale. This finding suggests that the effect of surfactant on droplets in homogeneous isotropic turbulence can be mainly summarised as a reduction in surface tension. The lack of a large-scale and time-persisting velocity difference among the carrier and dispersed phase, as found in up-flow and down-flow configurations \citep{LuMT_2017,takagi2008effects}, prevents the formation of significant, large-scale Marangoni stresses at the interface. Hence, in our simulation setup, Marangoni stresses play a minor, local role, with negligible effects on the statistics of the droplets. Local flow statistics better show the effect of these tangential stresses: Marangoni stresses modulate the flow at the interface by reducing the velocity component normal to the interface and increasing the tangential component. When computing the flow topology parameter at the interface, we find that Marangoni stresses increase the elongational component and reduce rotational flow at the interface. This result is coherent with the action of Marangoni stresses generating an elongational type of flow with sources corresponding to low-surface-tension regions and sinks to high-surface-tension regions. Inside the droplets, Marangoni stresses reduce elongational flow and increase the pure shear contribution.}

In conclusion, we find that a statistically homogeneous and isotropic flow allows for a simplified treatment of the surfactant effects. Results from \citeauthor{Hinze1955}'s theory can thus be applied to surfactant-laden flows, by considering the average surface tension reduction operated by the surfactant. For both clean and surfactant-laden flows, the Kolmogorov-Hinze scale separates two regimes characterised by very different scaling for the surface area of the droplets: a significant increase in the surface area is observed for droplets larger than the Kolmogorov-Hinze scale. This latter result has important implications for environmental and industrial multiphase flows, where the interface serves as a conduit for all species, momentum and energy transfers among the phases.

\section{Acknowledgements}
All authors acknowledge the computational resources provided on the Deigo cluster by the Scientific Computing and Data Analysis section of Research Support Division at OIST, and on 
Oakbridge-CX provided by the Information Technology Center (The University of Tokyo) through the HPCI System Research Project (Project ID: hp220100). I.C. would like to thank Julian Katzke (OIST) for guidance with the rendering software Blender, and David O'Connell (OIST) for helpful discussion of the Euler characteristic. 

\section{Funding}
The research was supported by the Okinawa Institute of Science and Technology Graduate University (OIST) with subsidy funding from the Cabinet Office, Government of Japan.

\section{Declaration of interests}
The authors report no conflict of interest.

\section{Data availability statement}
\ours{The plotted data are available at \url{https://groups.oist.jp/cffu/Cannon2024JFluidMech},} and the code used for calculating droplet morphology can be found at \url{https://github.com/marco-rosti/CFF-dropStats.git}.

\appendix

\section{Euler characteristic of a droplet interface}
\label{sec:Euler}
Here we use concepts from algebraic topology to derive equation~\ref{eq:handles}, which relates the Euler characteristic of a droplet interface to the number of voids and handles it contains. The Euler characteristic and genus are well-defined for a single surface; however, as seen in the image in figure~\ref{fig:genus}b, our droplets can have many distinct surfaces. For a general droplet with a number of voids $v$, it has an outer surface with Euler characteristic $\chi_0$, and the voids create $v$ distinct inner surfaces with Euler characteristics ${\chi_1, \chi_2,\ldots\chi_v}$. We define the Euler characteristic of the droplet's interface as the sum of the Euler characteristics of all the surfaces,
\begin{equation}
\chi=\sum_{i=0}^v\chi_i.
\label{eq:sumEuler}
\end{equation}
Both the inner and outer surfaces of the droplet can have handles. For any distinct orientable surface, the number of handles $h_i$ is its genus, and the genus is related to the Euler characteristic by $\chi_i=2-2h_i$ \citep[p.30]{massey_basic_1997}. Hence equation~\ref{eq:sumEuler} can be written and rearranged to
\begin{equation}
\chi=\sum_{i=0}^v (2-2h_i)=2(1+v)-\sum_{i=0}^v 2h_i \\
\end{equation}
The number of handles on the drop is the sum of the handles on all the droplet's surfaces, i.e., $h=\sum_{i=0}^v h_i$, so we can write,
\begin{equation}
\chi=2+2v-2h.
\end{equation}
This can be easily rearranged to obtain equation~\ref{eq:handles}.

\section{Validation of the surfactant model}
\label{sec:validation}
We report in this appendix the validation tests of the proposed numerical method. Two benchmark tests are presented and compared against existing theories, experiments and simulations.

\begin{figure}
  \centerline{
  \begin{tikzpicture}
    \node at (0,0) {\includegraphics[width=.55\textwidth]{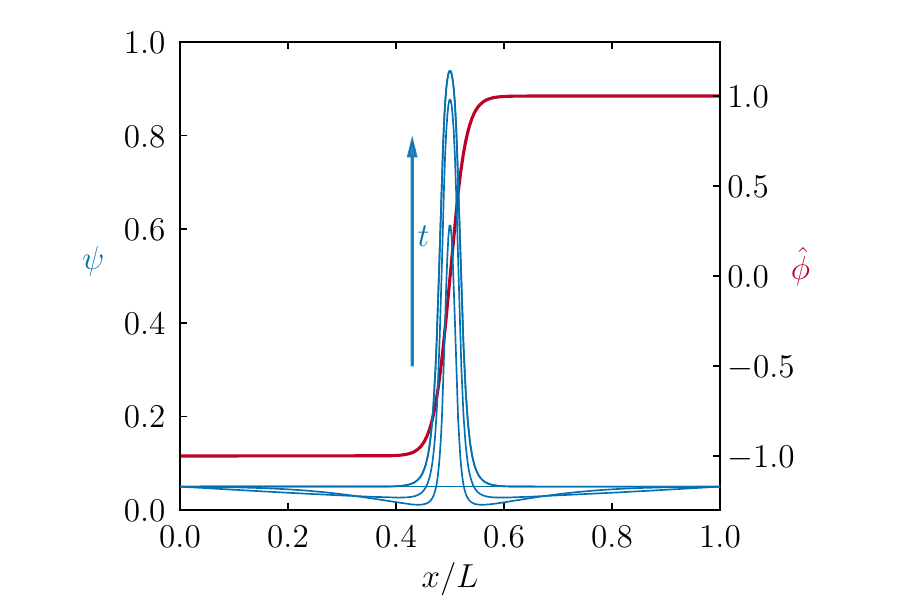}};
    \node at (7,0) {\includegraphics[width=.55\textwidth]{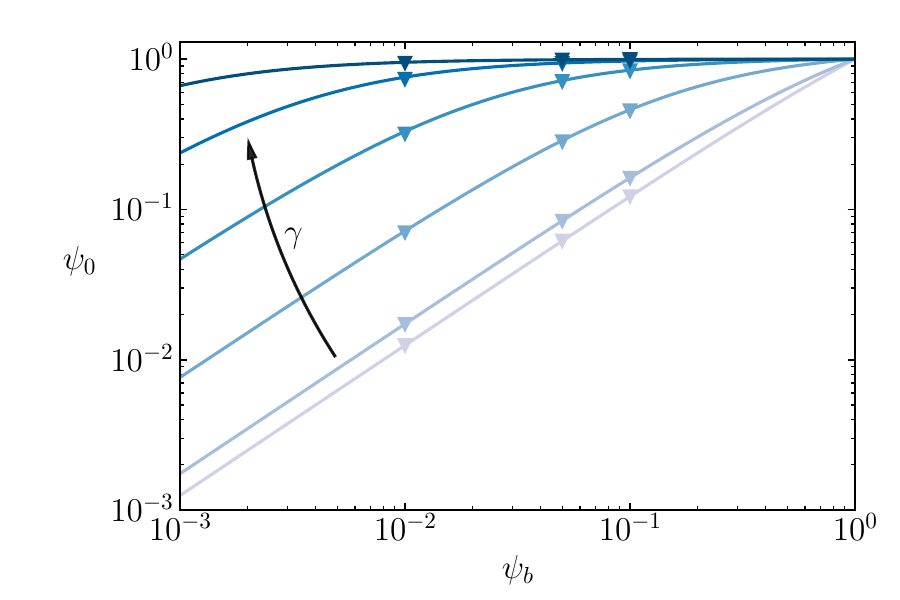}};
    \node at (-3.0,1.8){$(a)$};
    \node at (+4.0,1.8){$(b)$};
  \end{tikzpicture}
  }
  \caption{$(a)$~Adsorption of surfactant onto an interface at $x=L/2$. The right-hand axis shows the smoothed colour function $\hat\phi$ (in red), which has a smoothing width $3\Delta$. The left-hand axis shows the surfactant concentration (in blue) at four time-instants during the simulation.
  The bulk surfactant concentration is $\psi_b=0.05$ and the energy cost  of surfactant in the bulk is $\gamma=11.1\alpha$. $(b)$~The equilibrium interfacial surfactant concentration $\psi_0$ for a range of $\psi_b$ and $\gamma$. Lines show Langmuir isotherms, given by equation~\ref{eq:langmuir}, and markers show the results of our simulations. The value of $\gamma$ is represented by shading from light blue to dark blue.}
\label{fig:langmuir}
\end{figure}

We verify at first the adsorption dynamics of the surfactant at the interface: we start from a uniform surfactant concentration, equal to the surfactant concentration in the bulk $\psi_b$, and let the surfactant adsorb onto the interface (located at $x=L/2$, in figure~\ref{fig:langmuir}$a$). Initially, the system is out of equilibrium and surfactant diffuses towards the interface to restore the equilibrium; at equilibrium, the chemical potential is equal everywhere. 
We can thus equate the chemical potential in the bulk ($\hat\phi=\pm1$, surfactant concentration $\psi_b$) and at the interface ($\hat\phi=0$, surfactant concentration $\psi_0$) to obtain the so-called Langmuir isotherms \citep{Engblom2013}. For given values of the parameters of the chemical potential, $\alpha$, $\beta$ and $\gamma$, the Langmuir isotherm relates the surfactant concentration at the interface $\psi_0$ to that in the bulk $\psi_b$, in equilibrium conditions
\begin{equation}
\psi_0=\frac{\psi_b}{\psi_b+(1-\psi_b) e^{-\frac{\beta+\gamma}{2\alpha}}}.
\label{eq:langmuir}
\end{equation}
Figure~\ref{fig:langmuir}$a$ shows the setup we use to test our code against the Langmuir isotherm benchmark: a flat interface is located at $x=L/2$, and surfactant is initially uniformly distributed with a concentration equal to $\psi_b$. The flow is initially at rest and, due to the absence of any forcing, stays at rest throughout the entire simulation; similarly, the interface does not move from its initial position. At the boundaries $x=0$ and $x=L$, we impose a far-field value of surfactant, $\psi=\psi_b$, thus allowing surfactant to enter the system. In the $y$ and $z$ directions, we impose periodic boundary conditions. This benchmark is, in principle, a one-dimensional test ($x$ direction). However, we perform three-dimensional numerical simulations in order to use the very same solver described in section~\ref{sec:method}; all variables are uniform in the $y$ and $z$ directions. We resolve the flow and volume of fluid equations on a $N_x \times N_y \times N_z=100\times4\times4$ computational grid, and use a more refined grid for the surfactant transport equations, $500\times4\times4$. The smoothing width of the interface is $3\Delta$. We use $\beta/\alpha=0.741$ and explore a range of values for the energy cost in the bulk $\gamma/\alpha=\{0.0741,  0.741,  3.70,  7.41, 11.1, 14.8\}$. Figure~\ref{fig:langmuir}$b$ shows the simulated interfacial surfactant concentrations $\psi_0$ once they had reached equilibrium, for various values of $\psi_b$. The values obtained from our numerical simulations fall on top of the corresponding Langmuir isotherms, proving that the implemented numerical method can correctly capture the surfactant dynamics.

\begin{figure}
  \centerline{
  \begin{tikzpicture}
    \node at (0,0) {\includegraphics[width=.55\textwidth]{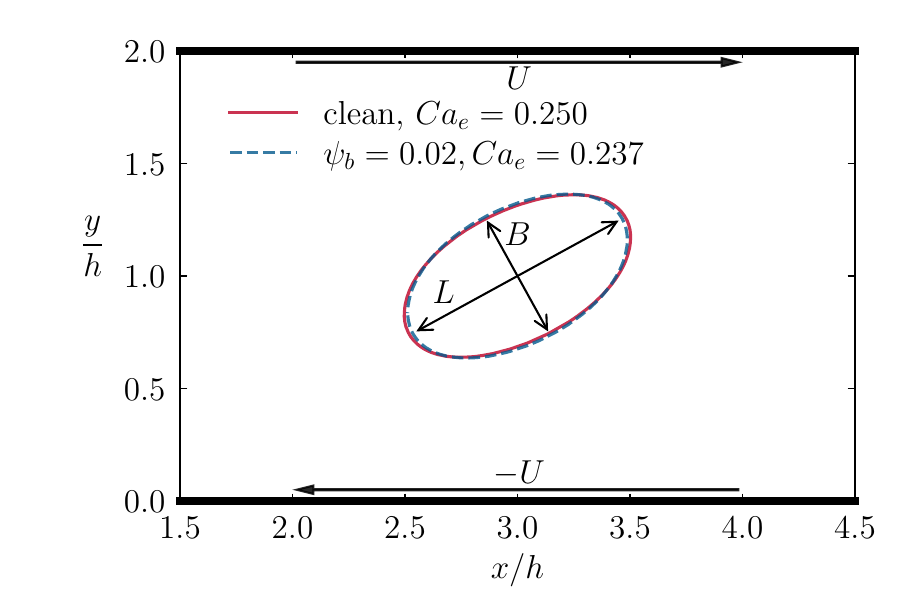}};
    \node at (7,0) {\includegraphics[width=.55\textwidth]{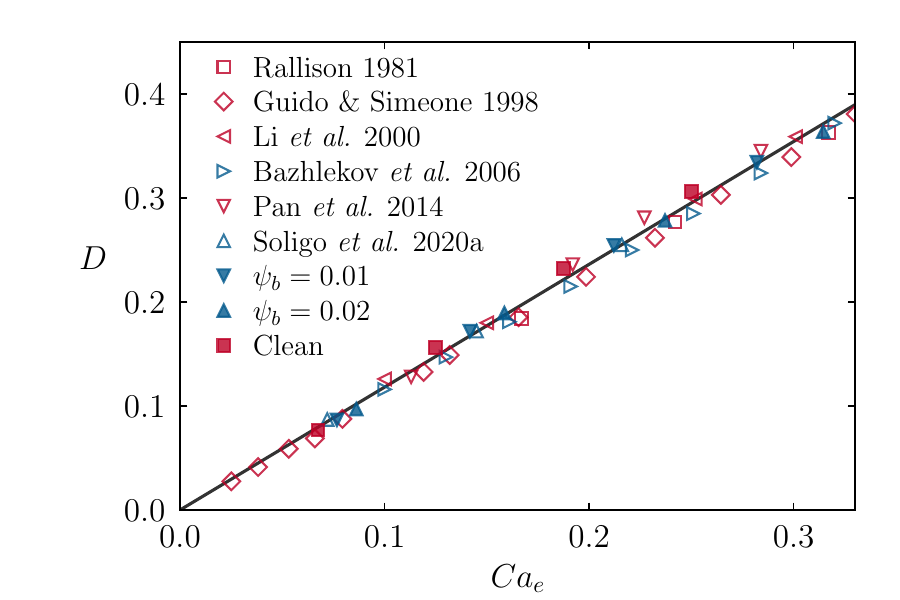}};
    \node at (-3.0,1.8){$(a)$};
    \node at (+4.0,1.8){$(b)$};
  \end{tikzpicture}
  }
  \caption{Deformation of droplets in shear flows. $(a)$~The simulated 2D domain (partially shown here) with velocity boundary conditions $u=U$ at $y=2h$, $u=-U$ at $y=0$, and periodic boundary conditions at $x=0$ and $x=6h$. We show the location of the droplet interface for a clean and surfactant-laden case. $(b)$~Dependence of the steady-state deformation parameter $D$ on the effective capillary number $Ca_e$. Our simulation results are shown with solid markers, and values from the literature are shown with empty markers. Clean droplets are marked in red, and surfactant-laden droplets are marked in blue. The solid line is the analytical relation from \citep{TAYLOR1934} with the confinement correction proposed by \cite{shapira1990}.}
\label{fig:taylor}
\end{figure}

The second benchmark aims to verify the surfactant transport on a moving interface and the computation of surface tension forces by measuring the deformation of a droplet in shear flow. 
\cite{TAYLOR1934} quantified the deformation of a droplet in a shear flow using the deformation parameter
\begin{equation}
D\equiv\frac{L-B}{L+B},
\end{equation}
where $L$ and $B$ are the droplet's largest and smallest principal diameters, respectively. Figure~\ref{fig:taylor}a shows the simulation setup that we use to reproduce Taylor's experiment. The effective capillary number $Ca_e\equiv a U \mu /h\langle\sigma\rangle$ describes the ratio of viscous forces to surface tension forces in the system, where $h$ is the domain half-height, $a=0.4h$ is the initial radius of the droplet, $\pm U$ is the fluid velocity of the top and bottom walls respectively, $\mu$ is the dynamic viscosity, which is the same for the bulk and droplet phases, and $\langle\sigma\rangle$ is the (average) surface tension at the droplet interface. As was previously demonstrated by \cite{soligo2020deformation}, there are negligible differences between the two-dimensional (2D) and three-dimensional (3D) cases in the limit of small Reynolds and Capillary numbers, as those considered here. Hence, we chose to perform two-dimensional numerical simulations to reduce the computational cost of the benchmark simulations. We select a computational grid with $N_x \times N_y \times N_z=600\times 200 \times 4$ points for the flow and volume of fluid variables, and we use a twice more refined grid to discretize the surfactant transport equation. As we use the same three-dimensional solver introduced in section~\ref{sec:method}, we have to use $4$ grid points in the $z$ direction; however, all variables are uniform in the $z$ direction.

We run a number of simulations with various values of $Ca_e$ and three values of the initial surfactant concentration in the bulk phase $\psi_b\in\{0,0.01,0.02\}$, where $\psi_b=0$ is the clean droplet case. In all cases, the Reynolds number of the flow $Re\equiv \rho U h/\mu$ is $Re=0.1$. The droplets are initially circular and deform in the shearing flow. Figure~\ref{fig:taylor}$b$ shows our measured values of the deformation parameter $D$ once they had reached a steady state. We compare our results (solid markers) with 2D droplets simulated using the boundary integral method \citep{rallison_numerical_1981}, experiments of 3D droplets \citep{guido1998}, 3D droplets simulated using the volume-of-fluid method \citep{li2000}, 3D droplets simulated using the boundary integral method with insoluble surfactants \citep{bazhlekov2006numerical}, 3D droplets simulated using dissipative particle dynamics \citep{pan_dissipative_2014}, and surfactant-laden droplets simulated using a phase-field method \citep{soligo2020deformation}. Our results are in good agreement with all of the above and also closely follow the analytical result from \citet{TAYLOR1934}, with the confinement correction proposed by \citet{shapira1990}, i.e., for droplets and carrier fluid having the same viscosity, the Taylor deformation parameter is equal to
\begin{equation}
D=\frac{35}{32} Ca_e\left[1+C_{S H} \frac{3.5}{2}\left(\frac{a}{2h}\right)^3\right],  
\end{equation}
where $C_{SH}=5.6996$ is a numerical coefficient accounting for the confinement due to the top and bottom boundaries \citep{shapira1990}.

In this appendix, we have shown that the proposed numerical method can accurately simulate the transport of surfactant over moving and deforming interfaces, \ours{as well} as the action of surfactant on surface tension. 


\end{document}